\pdfoutput=1
\documentclass[floatfix,reprint,aps,prc,showpacs,showkeys,eqsecnum,amsmath,amsfonts,amssymb,twocolumn,groupedaddress,superscriptaddress,nofootinbib,noeprint]{revtex4-2}

\usepackage{graphicx}    

\usepackage{float} 

\usepackage{dcolumn}     
\usepackage{bm}          
\usepackage[hyperindex,breaklinks]{hyperref}

\usepackage{amsmath} 
\usepackage{amssymb} 
\usepackage{amsfonts} 
\usepackage{latexsym}
\usepackage{enumitem} 

\usepackage{booktabs} 

\usepackage{cleveref}

\usepackage[english]{babel}
\usepackage[expansion=all,babel]{microtype}

\usepackage[usenames]{xcolor}

\usepackage{mathtools}

\usepackage{multirow}

\usepackage{rotating}

\newcommand{\sat}[1]{{#1}_{\mathrm{sat}}}
\newcommand{\sym}[1]{{#1}_{\mathrm{sym}}}
\newcommand{\asym}[1]{{#1}_{\mathrm{asym}}}
\newcommand{\eff}[1]{{#1}_{\mathrm{eff}}}
\newcommand{\cs}{c_\mathrm{s}}

\newcommand{\mn}{m_{\mathrm{eff;\,n}}^{\mathrm {PNM}}}
\newcommand{\mni}[1]{m_{\mathrm{eff;\,n;}\,#1}^{\mathrm{PNM}}}

\newcommand{\mN}{m_{\mathrm{eff;\,N}}^{\mathrm {SNM}}}
\newcommand{\mNi}[1]{m_{\mathrm{eff;\,N;}\,#1}^{\mathrm{SNM}}}

\usepackage{txfonts}

\newcommand{\be}{\begin{equation}}
\newcommand{\ee}{\end{equation}}
\newcommand{\ba}{\begin{eqnarray}}
\newcommand{\ea}{\end{eqnarray}}
   
\newcommand{\Msun}{M_\odot}

\makeatletter
	\newcommand{\vast}{\bBigg@{2.85}}
\makeatother

\newcommand{\eq}[1]{Eq.~\eqref{#1}}

\begin{document}

\preprint{Grant\#}

\author{Mikhail V. Beznogov}
\email{mikhail.beznogov@nipne.ro}
\affiliation{National Institute for Physics and Nuclear Engineering (IFIN-HH), RO-077125 Bucharest, Romania}

\title{Bayesian inference of the dense matter equation of state built upon extended Skyrme interactions}
\author{Adriana R. Raduta}
\email{araduta@nipne.ro}
\affiliation{National Institute for Physics and Nuclear Engineering (IFIN-HH), RO-077125 Bucharest, Romania}

\date{\today}

\begin{abstract}
    The non-relativistic model of nuclear matter with Brussels extended Skyrme interactions is employed in order to build, within a Bayesian approach, models for the dense matter equation of state (EOS). 
    In addition to a minimal set of constraints on nuclear empirical parameters; the density behavior of the energy per particle in pure neutron matter (PNM); a lower limit on the maximum neutron star (NS) mass, we require that the Fermi velocity of neutrons ($v_{\mathrm{F;\,n}}$) in PNM and symmetric nuclear matter (SNM) with densities up to $0.8~\mathrm{fm}^{-3}$ (arbitrary) does not exceed the speed of light. 
    The latter condition is imposed in order to cure a deficiency present in many Skyrme interactions [Duan and Urban, Phys. Rev. C 108, 025813 (2023)]. 
    We illustrate the importance of this constraint for the posterior distributions. 
    Some of our models are subjected to constraints on the density dependence of neutron (nucleon) Landau effective mass in PNM (SNM), too. 
    The impact of various sets of constraints on the behaviors of nuclear matter and NSs is discussed in detail.
    Systematic comparison with results previously obtained by employing Skyrme interactions is done for posteriors of both nuclear matter (NM) and NS parameters. Special attention is given to the model and constraints dependence of correlations among various quantities.
\end{abstract}

\maketitle

\section{Introduction}
\label{sec:Intro}

The collective properties of the dense and strongly interacting matter found in the interiors of neutron stars (NSs) depend upon the baryonic equation of state (EOS)~\cite{Oertel_RMP_2017,Burgio_PPNP_2021,Compose_EPJA_2022}.

In the past, NS EOSs have been obtained from EOSs of cold nuclear matter (NM) by solving the $\beta$-equilibrium equation, which determines the elementary composition of matter. 
Limited information extracted from nuclear structure experiments, which explore densities around the nuclear saturation density ($\sat{n} \approx 0.16~\mathrm{fm}^{-3} \approx 2.7 \times 10^{14}~\mathrm{g/cm^3}$) and small isospin asymmetries (i.e. with commensurate numbers of neutrons and protons), and from ab initio calculations of pure neutron matter (PNM) with densities up to $\sat{n}$~\cite{Wiringa_PRC_1988},
was reflected in a large scattering of model predictions at densities in excess of $\sim 2\sat{n}$.
The considerable efforts made by the heavy ion collision community improved the situation only marginally. The limited success of these attempts is primarily due to the large instrumental uncertainties that affect high multiplicity experiments~\cite{Andronic_PRC_2003,Reisdorf_NPA_2004,Reisdorf_NPA_2007,Zinyuk_PRC_2014,Russotto_PRC_2016}. Extra ambiguities stem from the model dependence of the results, use of simplified functionals of effective interactions in most molecular dynamics simulation codes and insufficient exploration of the parameter spaces~\cite{Zhang_PRC_2018,Ono_PRC_2019}.  

Multi-messenger observations of NSs, whose number and precision are growing fast, radically changed the situation.
Measurements of pulsars with masses around or larger than $2~\Msun$~\cite{Demorest_Nature_2010, Antoniadis2013, Arzoumanian_ApJSS_2018, Cromartie2020, Fonseca_2021} rekindled the debate on the emergence of non-nucleonic particle degrees of freedom, e.g., hyperons, mesons and quarks, in the cores of massive NSs~\cite{Oertel_JPG_2015,Chatterjee_EPJA_2016,Klahn_PLB_2007,Sedrakian_PPNP_2023}.
The interpretation of the outcome of the NSs coalescence in the GW170817 event~\cite{Abbott_PRL_2017} as the collapse of a hypermassive star into a black hole provided an upper boundary ($\lesssim 2.17\Msun$)~\cite{Margalit_ApJ_2017} for the maximum gravitational mass NSs can sustain and, thus, further challenged our understanding on the high density behavior of the NS EOS.
The measurement of the combined tidal deformability of NSs with masses $1.17 \lesssim M/\Msun \lesssim 1.60$ in the GW170817 event~\cite{Abbott_PRL_2017,Abbott_PRX_2019} supplied the first constraints on the intermediate density behavior of neutron rich matter~\cite{Abbott_PRL_121}.
Soon after that, the equatorial radius ($R_{\mathrm{eq}}=12.71^{+1.14}_{-1.19}~{\mathrm{km}}$~\cite{Riley_2019} or $R_{\mathrm{eq}}=13.02^{+1.24}_{-1.06}~{\mathrm{km}}$~\cite{Miller_2019}) and mass ($M=1.34^{+0.15}_{-0.16}\Msun$~\cite{Riley_2019} or $M=1.44^{+0.15}_{-0.14}\Msun$~\cite{Miller_2019}) of PSR J0030+0451 were estimated based on the analysis of the X-ray pulse profile measured by NICER and yielded extra information on the intermediate density behavior of the NS EOS.  
Knowledge on potentially broader domains of density and isospin asymmetry became later on available due to measurements of the radius of PSR J0740+6620 ($R_{\mathrm{eq}}=12.39^{+1.30}_{-0.98}~{\mathrm{km}}$~\cite{Riley_may2021} or $R_{\mathrm{eq}}=13.7^{+2.6}_{-1.5}~{\mathrm{km}}$~\cite{Miller_ApJL_2021}), whose mass is $2.08 \pm 0.07 \Msun$~\cite{Cromartie2020,Fonseca_2021}.
In addition to rotating hot spot patterns measured by NICER, the analysis of PSR J0740+6620~\cite{Riley_may2021,Miller_ApJL_2021} also benefited from X-ray observations by XMM-Newton.
All NICER figures quoted in this paragraph are at 68\% credible interval. 

The heterogeneity of constraints along with their dissimilar and not yet sufficiently well understood sensitivities to various domains of density and isospin asymmetry and complex shapes of credibility regions (CRs) commend for statistical analyses. 
The strategy consists in generating a large number of EOS models, that ideally explore the whole range of possibilities, and then ``filter" them based on how well they perform with respect to some criteria.
The criteria used so far belong to three classes:
i) theoretical calculations of NM,
ii) experimental nuclear physics data,
iii) astrophysical observations of NSs.
The multi-messenger probes of NSs interiors listed in the previous paragraph obviously fall in the third category.
Examples of constraints belonging to the first class are: the nuclear empirical parameters (NEPs; for their definition, see Sec.~\ref{sec:Model}), the results of various ab initio calculations of PNM with densities up to $\sim 1.1 \sat{n}$~\cite{Gandolfi_PRC_2012,Hebeler_ApJ_2013,Carlson_RMP_2015,Tews_ApJ_2018,Drischler_PRC_2020}, the results of perturbative quantum chromodynamics (pQCD) calculations~\cite{Kurkela_PRD_2010,Fraga_ApJL_2014}.
The latter of these criteria is the most challenging from the conceptual point of view. It consists in checking the compatibility between the behavior of EOSs at densities $n \lesssim 5\sat{n}$ and the behavior of matter in the quark phase, with densities $n \sim 40 \sat{n}$, by searching for causal and thermodynamically stable and consistent ``connections" of the two regimes~\cite{Komoltsev_PRL_2022,Gorda_ApJ_2023}.
Examples of constraints belonging to the second class are: neutron skin thicknesses, which include the recent PREX-II~\cite{PREX-II} and CREX~\cite{CREX} measurements of $^{208}$Pb and $^{48}$Ca, respectively; particle production~\cite{Reisdorf_NPA_2007,Zinyuk_PRC_2014,Russotto_PRC_2016} and flow~\cite{Andronic_PRC_2003,Russotto_PRC_2016} in relativistic heavy ion collisions.

Regarding the EOS models, two strategies have been adopted. The first one corresponds to the so-called schematic models and is by far the most popular. 
Parametric models, e.g., piece-wise polytropes, spectral parameterizations and (piece-wise) parameterizations of the speed of sound, and non-parametric models, e.g., Gaussian processes, stand out by their increased flexibility and low computational cost.
Yet, from the nuclear physics point of view, they are far from ideal.
The reasons are that they disregard all nuclear physics information that cannot be cast into a NS EOS format and, with the exception of a phase transition to quark matter, they completely ignore the particle composition of matter.
Inferences of NS EOSs based on schematic models have been obtained for different sets of constraints.
For instance, Refs.~\cite{Lynch_PLB_2022,Huth_Nature_2022,Koehn_2024} accounted for constraints from heavy ion collisions.
Refs.~\cite{Lynch_PLB_2022,Koehn_2024} carefully investigated the consequences of implementing constraints from PREX-II~\cite{PREX-II} and CREX~\cite{CREX} neutron skin measurements.
Refs.~\cite{Kurkela_ApJ_2014,Annala_PRL_2018,Lope_JPG_2019,Annala_Nature_2020,Annala_PRX_2022,Koehn_2024} implemented constraints from pQCD.
Refs.~\cite{Abbott_PRL_121,Landry_PRD_2019,Essick_PRD_2020,Raaijmakers_may2021} implemented exclusively constraints from astrophysical observations of NSs.
Refs.~\cite{Annala_PRL_2018,Huth_Nature_2022,Annala_PRX_2022,Koehn_2024} employed constraints from $\chi$EFT~\cite{Carlson_RMP_2015,Tews_ApJ_2018,Drischler_PRC_2020}.
By alternatively incorporating constraints belonging to different categories, Ref.~\cite{Koehn_2024} wonderfully demonstrates the role of each constraint.

The second strategy consists in using phenomenological, nuclear physics motivated, models.
The statistical inferences of the dense matter EOSs performed in the last years have employed a large collection of such models, e.g., Taylor expansion of the energy per nucleon with respect to the deviations from saturation and isospin symmetry \cite{Zhang_ApJ_2019,Ferreira_PRD_2020,Patra_PRD_2022}, semiagnostic metamodels \cite{Guven_PRC_2020}, a $\chi$EFT inspired expansion of the energy per nucleon in terms of neutrons and protons Fermi momenta \cite{Lim_EPJA_2019}, non-relativistic mean field models with various effective interactions~\cite{Papakonstantinou_2023,Beznogov_ApJ_2024,Imam_PRD_2024} and relativistic mean field models (RMF) with non-linear~\cite{Traversi_ApJ_2020,Malik_PRD_2023,Papakonstantinou_PRC_2023,Imam_PRD_2024} and density-dependent~\cite{Malik_ApJ_2022,Beznogov_PRC_2023,Char_PRD_2023} couplings.
The major advantage of phenomenological models lies in controlling the effective interactions and matter composition.
The former feature allows one to ``export" knowledge of the EOS in environments other than NSs, e.g., cold or hot matter with arbitrary charge fraction.
The obvious drawbacks are the limited flexibility, non-negligible computational costs and, in principle, spurious correlations among parameters.
As with the schematic models, different sets of constraints have been considered.
The ensemble of these studies made it possible to address the issue of correlations among properties of NM and properties of NSs and the sensitivity of posterior distribution upon implementation of different constraints.

The role played by the energy density functional in phenomenological models can be judged by comparing the results of Refs.~\cite{Malik_ApJ_2022,Malik_PRD_2023,Beznogov_PRC_2023} and Ref.~\cite{Beznogov_ApJ_2024}, which rely on relativistic and non-relativistic mean field models, respectively and employ the same {\em minimal set} of constraints.
It comes out that NS EOSs based on standard Skyrme interactions are softer than those built based on the RMF model with density dependent couplings, so that the NSs built within the first framework are characterized by values of radii and tidal deformability lower than those built within the second framework.
Significant differences manifest also in the neutron enrichment of the core. As a result, models in Ref.~\cite{Beznogov_ApJ_2024} are more prone to allow for direct Urca process than those in Refs.~\cite{Malik_ApJ_2022,Beznogov_PRC_2023}. 
A remarkable result of Ref.~\cite{Beznogov_ApJ_2024} is that accounting for the correlations between the values that the energy per particle $(E/A)$ in PNM takes at different densities not only brings further constraints into the isovector channel, but also couples it strongly to the isoscalar channel.
The impact of extra constraints from NS observations and heavy ion collisions was addressed in Ref.~\cite{Imam_PRD_2024}, which employed a RMF model with non-linear couplings and a non-relativistic mean field model with standard Skyrme interactions.
According to Ref.~\cite{Imam_PRD_2024}, accounting for astrophysical observations results in washing out much of the model-dependence of the posteriors of the NS-related parameters. Remarkably, NM posteriors and NS EOSs continue to be model dependent.

The aim of the present work is to pursue the Bayesian study of purely nucleonic dense matter via EOSs built within mean-field models of NM.  
Several avenues are taken to this aim. 
First, the Brussels extended Skyrme functional is employed for generating effective interactions.
Compared to standard Skyrme interactions, these extended interactions contain extra density dependent terms, which translates into an enhanced flexibility.
The first expected benefit of this enhanced flexibility is the possibility to account for a range of behaviors of NS EOSs wider than that allowed by the standard Skyrme interactions~\cite{Beznogov_ApJ_2024}. 
Second, this parametrization is expected to be able to provide for the density dependence of the Landau effective mass a behavior similar to the one predicted by ab initio calculations with three body forces~\cite{Burgio_PRC_2020,Drischler_PRC_2021}.
Then, in addition to a set of constraints we comment on in the following paragraph, we ask our models to comply with the requirement that up to a certain density the Fermi velocity of neutrons $v_{\mathrm{F;\,n}}$ in both PNM and symmetric nuclear matter (SNM) does not exceed the speed of light.  
This condition is introduced to cure the deficiency signaled recently by Duan and Urban~\cite{Urban_PRC_2023}.
It affects many of the widely used Skyrme parametrization and, thus, makes them unphysical.
The value of the density up to which we enforce the condition $v_{\mathrm{F;\,n}}<c$ is, in most cases, $0.8~\mathrm{fm}^{-3}$ (arbitrarily chosen). Other values are considered for illustrative purposes, too. 

With the exception of the condition for $v_{\mathrm {F;\,n}}$, which is new, sets of constraints equivalent to those used in Refs.~\cite{Malik_ApJ_2022,Malik_PRD_2023,Beznogov_PRC_2023,Beznogov_ApJ_2024} are employed throughout this paper.
As such, the present results can be directly compared with those obtained within other widely used mean field models. 
This is a basic requirement for judging the model dependence of the conclusions.
The nuclear physics constraints we implement correspond to:
  i) the four best known NEPs, i.e., $\sat{n}$, $\sat{E}$, $\sat{K}$ and $\sym{J}$,
  ii) the density behaviors of $E/A$ and effective neutron mass in PNM ($\mn$) up to $0.16~\mathrm{fm}^{-3}$, 
  iii) the density dependence of the effective nucleon mass in SNM ($\mN$) up to $0.16~\mathrm{fm}^{-3}$. 
For the latter three quantities we utilize the results of $\chi$EFT calculations from Ref.~\cite{Drischler_PRC_2021}.
The only NS-like constraint we implement is the lower bound on the maximum gravitational mass of static NSs.
In most cases, the conservative value of $2~\Msun$ is used. 
For illustrative purposes, larger values will be considered in specific situations, too.
All EOS models we build are causal up to the density corresponding to the maximum mass configuration.
We also insure that they are thermodynamically stable.

The choice not to implement constraints on the tidal deformability of the intermediate mass NSs, extracted from GW170817~\cite{Abbott_PRL_2017,Abbott_PRX_2019}, and radii of intermediate and massive NSs, extracted from X-ray observations of PSR J0030+0451~\cite{Miller_2019,Riley_2019} and PSR J0740+6620~\cite{Miller_ApJL_2021,Riley_may2021}, respectively, can be viewed both as a drawback and an advantage.
The drawback refers to an insufficient constraining of the EOS over the density domain $\sat{n} \lesssim n \lesssim 3 \sat{n}$.
The advantage lies in keeping the results free of the systematic uncertainties that might affect those very sophisticated inference procedures, e.g., models of atmosphere, instrument response, geometry and temperature distributions of hot spots, sampling techniques, prior assumptions on the EOSs, etc.
Radii measurements of PSR J0030+0451 in Refs.~\cite{Miller_2019,Riley_2019} show that the result depends on the number, morphology and topology of hot spots that generate the observed pulse profile.  
The updated analysis of PSR J0030+0451 in Ref.~\cite{Vinciguerra_ApJ_2024} has shown that noise setting and technical aspects of the inference procedure also affect NICER mass-radius estimates and that the posterior surface has a multimodal structure.
For PSR J0740+6620~\cite{Miller_ApJL_2021,Riley_may2021}, separate choices for the relative area of XMM-Newton to NICER also contribute to differences in the results.
The situation with tidal deformabilities constraints extracted from GW170817 is roughly similar.
Ref.~\cite{Abbott_PRL_2017} comments that the inferred values depend on whether the same EOS was used for both NSs or not. 
Ref.~\cite{Abbott_PRX_2019} shows that the constraints on combined tidal deformability also depend on a number of assumptions, including the waveform model and NSs' spins. 
Even if our models are not conditioned on data extracted from NICER and LIGO-Virgo measurements, compliance with the constraints contributed by astrophysicists colleagues is thoroughly checked on a posteriori distributions.

The rest of the paper is organized as follows. Sec.~\ref{sec:Model} briefly reviews the theoretical approach.
The Bayesian setup is discussed in Sec.~\ref{sec:Bayes}.
Sec.~\ref{sec:Results} addresses the behaviors of NM and NS matter as well as possible correlations among NEPs or among NEPs and global parameters of NSs. 
Comparison with results obtained in Ref.~\cite{Beznogov_ApJ_2024}, which employed standard Skyrme interactions and similar sets of constraints, highlights the sensitivity of the results, correlations included, on the structure of the effective interaction and specific constraints.
The conclusions are drawn in Sec.~\ref{sec:Concl}. 
Ref.~\cite{Beznogov_ApJ_2024} will be hereafter referred to as Paper~I.

\section{The model}
\label{sec:Model}

In this paper, we use the Brussels extended Skyrme interactions~\cite{Chamel_PRC_2009}, which are obtained by supplementing the standard Skyrme interaction~\cite{Negele_PRC_1972} 
\begin{align}
	V\left({\mathbf r}_1, {\mathbf r}_2 \right) &= t_0 \left(1+x_0 P_{\sigma} \right) \delta\left( {\mathbf r}\right) \nonumber \\
	&+ \frac{t_1}2  \left(1+x_1 P_{\sigma} \right) \left[{\mathbf k}'^2 \delta\left( {\mathbf r}\right) +\delta\left( {\mathbf r}\right) {\mathbf k}^2 \right] \nonumber \\
    &+ t_2 \left(1+x_2 P_{\sigma} \right) {\mathbf k}' \cdot \delta\left( {\mathbf r}\right) {\mathbf k}
    \label{eq:VSk}\\
	&+ \frac{t_3}6 \left(1+x_3 P_{\sigma} \right) \left[n \left({\mathbf R} \right)\right]^{\sigma} \delta\left( {\mathbf r}\right) \nonumber \\
  	&+ i W_0 \left( \boldsymbol{\sigma}_1 + \boldsymbol{\sigma}_2\right) \cdot \left[ {\mathbf k}' \times \delta\left({\mathbf r}\right) {\mathbf k} \right], \nonumber
\end{align}
with two density dependent terms,
\begin{align}
    & V^{\mathrm{BSk}}\left({\mathbf r}_1, {\mathbf r}_2 \right) = V\left({\mathbf r}_1, {\mathbf r}_2 \right) \nonumber \\
    &+ \frac{t_4}2 \left(1+x_4 P_{\sigma} \right) \left[{\mathbf k}'^2 \left[n \left({\mathbf R} \right)\right]^{\beta} \delta\left( {\mathbf r}\right) + \delta\left( {\mathbf r}\right)  \left[n \left({\mathbf R} \right)\right]^{\beta} {\mathbf k}^2 \right]
    \label{eq:VBSk} \\
    &+ t_5 \left(1+x_5 P_{\sigma} \right){\mathbf k}'\cdot \left[n \left({\mathbf R} \right)\right]^{\gamma} \delta\left( {\mathbf r}\right) {\mathbf k}~. \nonumber
\end{align}

In Eqs. \eqref{eq:VSk} and \eqref{eq:VBSk} the following notations are used:
${\mathbf r}={\mathbf r}_1-{\mathbf r}_2$, ${\mathbf R}=\left({\mathbf r}_1+{\mathbf r}_2\right)/2$; ${\mathbf k}=\left(\boldsymbol{\nabla}_1- \boldsymbol{\nabla}_2\right)/2i$ is the relative momentum operator acting on the right and ${\mathbf k}'$ is its conjugate acting on the left; $P_{\sigma}=\left( 1+ \boldsymbol{\sigma}_1 \cdot \boldsymbol{\sigma}_2\right)/2$ is the two body spin-exchange operator; $n({\mathbf r})=n_\mathrm{n}({\mathbf r})+n_\mathrm{p}({\mathbf r})$ is the total local density; $n_i({\mathbf r})$ with $i=\mathrm{n},\mathrm{p}$ are the neutron and proton local densities. 

In the absence of spin polarization, the energy density of homogeneous matter with no Coulomb interaction is a sum of four terms
\begin{equation}
	{\cal H}=k+h_0+h_3+\eff{h},
	\label{eq:h}
\end{equation}
where $k$ is the kinetic energy term, $h_0$ is a density-independent two-body term, $h_3$ is a density-dependent term and $\eff{h}$ is a momentum-dependent term.
Each of these can be expressed in terms of densities of particles and kinetic energies as
\begin{align}
	&k = \frac{\hbar^2}{2m} \tau,
	\label{eq:k} \\
	&h_0 = C_0 n^2+D_0 n_3^2,
	\label{eq:h0} \\
	&h_3 = C_3 n^{\sigma+2}+D_3 n^{\sigma} n_3^2,
	\label{eq:h3} \\
	&\eff{h} = \eff{\widetilde C} n \tau+\eff{\widetilde D} n_3 \tau_3,
	\label{eq:heff}
\end{align}
where $n=n_\mathrm{n}+n_\mathrm{p}$ and $n_3=n_\mathrm{n}-n_\mathrm{p}$ stand for the isoscalar and isovector particle number densities; $\tau=\tau_\mathrm{n}+\tau_\mathrm{p}$ and $\tau_3=\tau_\mathrm{n}-\tau_\mathrm{p}$ denote the isoscalar and isovector densities of kinetic energy; $2/m=1/m_\mathrm{n}+1/m_\mathrm{p}$, where $m_i$ with $i=\mathrm{n},\mathrm{p}$ denotes the bare mass of nucleons. 

The coefficients $C_0$, $D_0$, $C_3$ and $D_3$ have the same expressions as for the standard Skyrme interaction~\citep{Ducoin_NPA_2006},
\begin{align}
    \begin{split}
	   &C_0 =  3t_0/8, \quad D_0 = -t_0\left( 2 x_0+1\right)/8, \\
	   &C_3 =  t_3/16, \quad D_3 = -t_3 \left( 2 x_3+1\right)/48,
    \end{split}
   \label{eq:C0D0}     
\end{align}
while the contributions of extra terms enter into $\eff{\widetilde C}$ and $\eff{\widetilde D}$
\begin{align}
    \begin{split}
	   &\eff{\widetilde C} = \eff{C} + \left[ 3 t_4 n^{\beta} +t_5 \left( 4 x_5+5\right) n^{\gamma}\right]/16, \\
	   &\eff{\widetilde D} = \eff{D} +\left[ -t_4 \left(2 x_4+1 \right) n^{\beta} +t_5  \left( 2 x_5+1\right) n^{\gamma}\right]/16,
    \end{split}
	\label{eq:CeffDeff_B}
\end{align}
where
\begin{align}
    \begin{split}
        &\eff{C}=\left[ 3 t_1+t_2 \left( 4 x_2+5\right)\right]/16, \\
        &\eff{D}=\left[t_2 \left( 2 x_2+1\right) -t_1 \left(2 x_1+1 \right)\right]/16,
    \end{split}  
  \label{eq:CeffDeff}
\end{align}
are the coefficients of momentum-dependent terms in the standard Skyrme interaction~\citep{Ducoin_NPA_2006}. 
With this convention there are 13 parameters ($C_0$, $D_0$, $C_3$, $D_3$, $\eff{C}$, $\eff{D}$, $t_4$, $x_4$, $t_5$, $x_5$, $\sigma$, $\beta$ and $\gamma$) that define the most general form of Brussels extended Skyrme parameterizations. 
In practice, the six parameters that enter Eq.~\eqref{eq:VBSk} are given fixed values~\cite{Chamel_PRC_2009,Goriely_PRC_2013,Goriely_NPA_2015}. 
The advantages of doing so are in controlling the density dependence of the extra momentum dependent terms and keeping the $\chi^2$ minimization procedure upon which the values of other parameters are obtained computationally cheap. 

A peculiarity of the zero-range Skyrme interactions, where the momentum-dependent part of the potential is proportional to ${\mathbf k}^2$, consists in the possibility of introducing effective masses
\begin{equation}
	\frac{\hbar^2}{2 m_{\mathrm{eff};\,i}}=\frac{\hbar^2}{2 m_i}+\eff{\widetilde C} n \pm \eff{\widetilde D} n_3,
	\label{eq:meff}
\end{equation}
that depend exclusively on particle densities; here ``+'' and ``--'' signs correspond to neutrons and protons, respectively.
In the limit of zero temperature, $m_{\mathrm{eff};\,i}$ correspond to the Landau effective masses $m_{\mathrm{Landau,\, eff};\,i}$, defined in terms of single particle density of states $de_i/dk_i$ at the Fermi surface,
\begin{equation}
   m_{\mathrm{Landau,\,eff};\,i}^{-1}=\frac1{k_i}  \left.\frac{d e_i}{d k_i}\right|_{k=k_{\mathrm{F};\,i}}.
\end{equation}

The presence of extra terms that depend on $t_4$, $x_4$, $\beta$, $t_5$, $x_5$, $\gamma$ is essential for alleviating the decrease of $m_{\mathrm{eff};\,i}$ with density. 
As such, two major deficiencies can be healed. 
The first one regards the neutron Fermi velocity
\begin{equation}
    v_{\mathrm{F;\,n}} = \frac{k_\mathrm{F;\,n}}{m_{\mathrm{eff;\,n}}}
    \label{eq:vF}
\end{equation}
that in dense matter might exceed the speed of light. 
Duan and Urban~\cite{Urban_PRC_2023} have recently shown that this is a general issue with standard Skyrme interactions, which obviously makes them unsuitable for describing NS matter.
Correlated with the previous aspect, the extra flexibility granted by the new terms can be exploited in order to achieve a $m_{\mathrm{eff};\,i}(n)$ behavior compliant with the one predicted by microscopic models.
Brueckner-Hartree-Fock calculations~\cite{Baldo_PRC_2014,Burgio_PRC_2020} performed up to the density $n=0.8~\mathrm{fm}^{-3}$ showed that three body (3N) forces are responsible for an U-shape behavior of $m_{\mathrm{eff};\,i}(n)$. 
The minimum value of $m_{\mathrm{eff};\,i}/m_{\mathrm N}$ of the order of $0.7 $ was obtained for densities $n$ ranging between $0.25~\mathrm{fm}^{-3}$ and $0.6~\mathrm{fm}^{-3}$, depending on the strength of 3N forces and isospin asymmetry. 
$\chi$EFT calculations~\cite{Drischler_PRC_2021} validate the qualitative findings of Refs.~\cite{Baldo_PRC_2014,Burgio_PRC_2020}, but predict that the minimum occurs at densities around $\sat{n}$.
We also note that the extra momentum dependent terms lift the degeneracy that standard Skyrme interactions manifest
for $\sigma=2/3$.

The behavior of NM at arbitrary values of $n$ and $\delta$ is commonly discussed in terms of NEPs.
The first set of NEPs corresponds to the coefficients in the Taylor expansion of the energy per particle $E/A={\cal H}/n = e/n$ in terms of the deviation $\chi=\left(n-\sat{n}^{\delta} \right)/3\sat{n}^{\delta}$ from the saturation density $\sat{n}^{\delta}$,
\begin{equation}
	E\left(n,\delta\right)/A=\sum_{i=0,1,2,...} \frac1{i!} \sat{X}^{\delta;\,i}\, \mathcal{X}^i,
	\label{eq:Taylor_chi}
\end{equation}
with
\begin{equation}
	\sat{X}^{\delta;\,i}=3^i \left(\sat{n}^{\delta}\right)^i \left. \left( \frac{\partial^i (e/n)}{\partial n^i}\right) \right|_{n=\sat{n}^{\delta}}.
\end{equation}
The (approximate) isospin invariance of the nucleon-nucleon interaction and the fact that SNM is more stable than isospin asymmetric matter allow for an alternative Taylor expansion, this time in terms of $\delta^2$~\citep{Chen_PRC_2009},
\begin{equation}
	E\left(n,\delta\right)/A=E_0(n,0)+\delta^2 E_{\mathrm{sym};\,2}(n,0)+\delta^4 E_{\mathrm{sym};\,4}(n,0) + \cdots
	\label{eq:Taylor_delta}
\end{equation}
Here, $E_0(n,0)$ corresponds to the saturation density of SNM and $E_{\mathrm{sym};\,k}$ stand for different order terms of the symmetry energy. Each of these can be further expanded in terms of the deviation $x=\left(n-\sat{n}^{0} \right)/3 \sat{n}^{0}$ from the saturation density of SNM,
\begin{equation}
	E_0(n,0)=\sum_{i=0,1,2,...} \frac1{i!} \sat{X}^{0;\,i} x^i,
\end{equation}
\begin{equation}
  E_{\mathrm{sym};\,k}(n,0)=\sum_{j=0,1,2,...} \frac1{j!} X_{\mathrm{sym};\,k}^{0;\,j} x^j,~ k=2,4,...
  \label{eq:Esymk}
\end{equation}
where $X_{\mathrm{sym};\,k}^{0;\,j}= \left. \left(\partial^j E_{\mathrm{sym};\,k}(n,0)/\partial x^{j} \right) \right|_{n=\sat{n}^0}$.

For asymmetric NM, the lowest order term in $\delta^2$ in Eq. \eqref{eq:Taylor_delta}, called the symmetry energy, and the per-nucleon cost of converting SNM in PNM, called asymmetry energy, are of major interest. 
Their expressions are
\begin{equation}
E_{\mathrm{sym};\,2}(n)= \frac12 \left. \frac{ \partial^2 \left(e/n \right)}{\partial \delta^2}\right|_{n,\,\delta=0},
\end{equation}
and
\begin{equation}
	E_{\mathrm{asym}}(n)=E(n, \delta=1)/A-E(n, \delta=0)/A,
	\label{eq:Easym}
\end{equation}
respectively.

The expressions of most frequently used NEPs are provided in Appendix~\ref{App:NEPs}.

\section{The Bayesian setup}
\label{sec:Bayes}

In this section, we discuss the various constraints imposed on our EOS models and the procedure adopted to explore the parameter space.


\subsection{Constraints}
\label{ssec:Constraints}

\renewcommand{\arraystretch}{1.20}
\setlength{\tabcolsep}{10pt}
\begin{table}
    \centering
    \caption{Constraints imposed on EOS models.
    $\sat{E}$ and $\sat{K}$ represent the energy per particle and compression modulus of symmetric saturated matter with the density $\sat{n}$; 
    $\sym{J} \equiv E_{\mathrm{sym};\,2}(\sat{n},0)$ stays for the symmetry energy at saturation;
    $\left(E/A\right)_i$ with $i=2,3,4$ stand for the energy per particle of PNM at the densities of 0.08, 0.12 and 0.16 $\mathrm{fm}^{-3}$;
    $\mni{i}$ ($\mNi{i}$) denotes the Landau effective mass of the neutron (nucleon) in PNM (SNM) at the density $n_i$;
    $M_{\mathrm{G}}^*$ represents the maximum gravitational mass of a NS.
    For all quantities except $M_{\mathrm{G}}^*$ we provide the mean and its standard deviation;
    for $M_{\mathrm{G}}^*$, we specify the threshold values.
    $m_\mathrm{n}$ and $m_\mathrm{N}$ represent the neutron and nucleon bare masses, respectively.}
    \begin{tabular*}{\columnwidth}{ccccc}
        \toprule
        \toprule
        Quantity             & Units              & Value   & Std. deviation & Ref. \\
        \midrule
        $n_{\mathrm{sat}}$   & $\mathrm{fm}^{-3}$ & 0.16 	& 0.004         & 1 \\
        $E_{\mathrm{sat}}$   & MeV 		          & $-15.9$ & 0.2 	        & 1 \\
        $K_{\mathrm{sat}}$   & MeV 		          & 240 	& 30 	        & 1 \\ 
        $J_{\mathrm{sym}}$   & MeV 		          & 30.8   & 1.6            & 1  \\
        $\left(E/A\right)_2$ & MeV 	              & 9.212  &  0.226         & 2 \\
        $\left(E/A\right)_3$ & MeV 	              & 12.356 &  0.512         & 2 \\
        $\left(E/A\right)_4$ & MeV 	              & 15.877 &  0.872         & 2 \\       
        $\mNi{2}$            & $m_\mathrm{N}$     & 0.715  &  0.011         & 2 \\       
        $\mNi{3}$            & $m_\mathrm{N}$     & 0.667  &  0.012         & 2 \\       
        $\mNi{4}$            & $m_\mathrm{N}$     & 0.638  &  0.013         & 2 \\       
        $\mni{2}$            & $m_\mathrm{n}$     & 0.877  &  0.004         & 2 \\
        $\mni{3}$            & $m_\mathrm{n}$     & 0.866  &  0.011         & 2 \\
        $\mni{4}$            & $m_\mathrm{n}$     & 0.880  &  0.026         & 2 \\
        $M_{\mathrm{G}}^*$   & $\Msun$ 	          & $>2.0$  &  ---          & 3  \\
        \bottomrule
        \bottomrule
    \end{tabular*}
    \label{tab:constraints}
    {\raggedright \textbf{References.} (1) \citet{Margueron_PRC_2018a}; (2) \citet{Drischler_PRC_2021}; (3) \citet{Fonseca_2021}.\par}
\end{table}
\setlength{\tabcolsep}{2.0pt}
\renewcommand{\arraystretch}{1.0}

\renewcommand{\arraystretch}{1.05}
\setlength{\tabcolsep}{3.0pt}
\begin{table}
  \centering
  \caption{Synoptic view of constraints on $(E/A)_i$, $\mNi{i}$ and $\mni{i}$ with $i=2,3,4$ and Fermi velocity ($v_{\mathrm F}$) whose implementation changes from one run to the others.}
  \begin{tabular}{lccccccc}
    \toprule
    \toprule
    Run   & $\left(E/A\right)_i$ &  correl.   & $\mNi{i}$ & correl. & $\mni{i}$ &  correl. & $v_{\mathrm{F;\,n}}$ \\
    \midrule
    0     &  \checkmark          & \checkmark &   --       &  --      & --         &  --    & \checkmark \\
    1     &  \checkmark          & --         &   --       &  --      & --         &  --    & \checkmark \\
    1$^*$ &  \checkmark          & --         &   --       &  --      & --         &  --    &  -- \\
    2     &  \checkmark          & --         & \checkmark &  --      & \checkmark & --     & \checkmark \\
    3     &  \checkmark          & --         & \checkmark &  --      & \checkmark & \checkmark & \checkmark \\
    \bottomrule
    \bottomrule
  \end{tabular}
  \label{tab:runs}
\end{table}
\setlength{\tabcolsep}{2.0pt}
\renewcommand{\arraystretch}{1.0}

As in our previous works~\cite{Beznogov_PRC_2023,Beznogov_ApJ_2024}, our strategy is to employ a minimum number of well-motivated constraints coming from nuclear physics data, $\chi$EFT calculations of NM with densities up to $\sat{n}$ and extreme isospin asymmetries and astrophysical observations of NSs. 
As such, quantities not yet constrained will span wide ranges, which is essential for potential correlations to manifest.

The constraints from nuclear physics data correspond to properties of NM close to saturation and are the following:
saturation density of SNM ($\sat{n}$), energy per particle of saturated SNM ($\sat{E}$), compression modulus of saturated SNM ($\sat{K}$), and the symmetry energy ($\sym{J}$) at $\sat{n}$.
Their values, listed in Table~\ref{tab:constraints}, are those calculated by Margueron~et~al.~\cite{Margueron_PRC_2018a}, who considered a collection of 35 Skyrme interactions that are frequently employed in the literature. 

\looseness=-1
The density behavior of PNM is constrained by the $\chi$EFT data from Ref.~\cite{Drischler_PRC_2021}, where nucleon-nucleon (NN) interactions computed at N$^3$LO are supplemented by three nucleon (3N) interactions computed at N$^2$LO.
All our runs account for $\chi$EFT predictions for $E/A$ in PNM, which are sufficient for constraining the low density behavior of the isovector channel. 
Some of our runs also account for $\chi$EFT results regarding the density behavior of $\mn$ and $\mN$.
The former is expected to further constrain the isovector channel, while it remains to be seen whether the latter adds to stronger constraints in the isoscalar sector or, on the contrary, reveals tension with the values adopted for NM in the vicinity of $(\sat{n},\delta=0)$. 
For all the quantities for which we use the predictions of $\chi$EFT~\cite{Drischler_PRC_2021}, constraints are imposed at three densities $n=0.08, 0.12, 0.16~\mathrm{fm}^{-3}$. 
The availability of the results corresponding to six Hamiltonians in Ref.~\cite{Drischler_PRC_2021} makes it possible to account for correlations among the values that each of these quantities takes at different densities, which strengthen the constraints~\cite{Beznogov_ApJ_2024}. 
For the means and standard deviations (SDs) of $(E/A)_i$, $\mni{i}$, $\mNi{i}$, see Table~\ref{tab:constraints}. 
For the different manner in which these constraints are implemented, see Table~\ref{tab:runs}. 
We mention that in addition to the runs in Table~\ref{tab:runs}, we also considered the case analogous to run~3 but with correlations among $\mNi{i}$ instead of correlations among $\mni{i}$. 
As the results of this run were very similar to the results of run~2, they are not included in the following figures. 
Nevertheless, some comments will be offered in the end of Sec.~\ref{sec:Results}.

\looseness=-1
Two extra constraints are added to both SNM and PNM to ensure that the models we generate are physical.
Specifically, we ask that in SNM and PNM 
i) for densities $n \leq 0.8~\mathrm{fm}^{-3}$, $0 \leq {\eff{m}}_{\!,\,i} / m_{\mathrm N} \leq 1$ with $i=\mathrm{n,p}$, 
ii) unless otherwise explicitly mentioned, for densities $n \leq 0.8~\mathrm{fm}^{-3}$ the neutron Fermi velocity ($v_{\mathrm{F;\,n}}$) does not exceed the speed of light. 
The threshold density of  $0.8~\mathrm{fm}^{-3}$ was chosen somewhat arbitrarily as a compromise between extending the validity domain of the model and computational efficiency.
It goes without saying that for $n>0.8~\mathrm{fm}^{-3}$, ${\eff{m}}_{\!,\,i} / m_{\mathrm N}$ and/or $v_{\mathrm{F;\,n}}/c$ may exceed 1, which implies that the EOS models we generate behave well over a density domain narrower than the one that is explored within the NSs built upon these models.
To better judge the implications of the condition on $v_{\mathrm{F;\,n}}$, a case where this constraint is disregarded is also considered.
More precisely, run~1$^*$ is a full analogue of run~1, but without the Fermi velocity constraint.
Results obtained when replacing the value of $0.8~\mathrm{fm}^{-3}$ with the values of $0.5~\mathrm{fm}^{-3}$ and $1.0~\mathrm{fm}^{-3}$ are considered too, see Appendix~\ref{App:vF}.

In addition, NS EOSs are required to be 
i) stiff enough to produce maximum NS masses in excess of $2~\Msun$, 
ii) thermodynamically stable ($P > 0$ and $dP/dn \geq 0$), 
iii) causal up to a density equal to the central density of the maximum mass configuration. 
The consequences of using a larger value for the lower limit of the maximum gravitational mass ($M_{\mathrm{G}}^* \geq 2.2 \Msun$) have been investigated as well; the conclusions are reported in the end of Secs.~\ref{sss:NEPs} and \ref{sss:NS}.

We note that usage of constraints similar to those implemented in Paper~I, where standard Skyrme interactions have been employed, allows to reveal the model-dependence of the results. 
We are particularly interested in the role played by the extra momentum dependent terms of the extended interaction, on the one hand, and the condition on the Fermi velocity, on the other hand. 
As a matter of fact, posterior distributions of various sets of models built here will be systematically confronted with some of the posteriors from Paper~I. 
We also note that a similar exercise, where results of standard Skyrme interactions were confronted with those of an RMF model with density dependent couplings~\cite{Beznogov_PRC_2023}, has been done in Paper~I. 

\vspace{-0.2cm}
\subsection{Likelihood}
\label{ssec:Likelihood}

From the likelihood point of view, the quantities on which we impose conditions can be classified into three categories: uncorrelated, correlated and threshold ones. 
In this paper, NEPs are always considered as belonging to the first category. 
Depending on whether correlations between the values that $\left( E/A \right)_i$; $\mni{i}$; $\mNi{i}$ take at different densities are accounted for or not, these quantities enter the second or the first category. 
Quantities whose values under specific conditions are used in order to accept or reject models proposed during the Monte Carlo exploration of the parameter space enter into the third category. 
Examples for this case are offered by the maximum NS mass, speed of sound, and Fermi velocity.
Effective masses in PNM and SNM enter into the third category as well, but, depending on the run, they might also enter into the first and/or the second categories.

The log-likelihood function of the run $q$ can be decomposed into three parts
\begin{align}
\log \mathcal{L}_{q}=\log \mathcal{L}_{q,\mathrm{uncorr.}} + \log \mathcal{L}_{q,\mathrm{corr.}}+ \log \mathcal{L}_{q,\mathrm{thresh.}}.
	\label{eq:Chi2}
\end{align}
The log-likelihood function for the uncorrelated constraints reads 
\begin{align}
	\log \mathcal{L}_{q,\mathrm{uncorr.}} \propto -\chi_{q,\mathrm{uncorr.}}^2 = -\frac{1}{2} \sum_{i=1}^{N_q^\mathrm{uncorr.}}  \left(\frac{d_i - \xi_i(\mathbf{\Theta}) } {\mathcal{Z}_i} \right)^2,
	\label{eq:Chi2-NoCorr}
\end{align}
where $N_q^\mathrm{uncorr.}$ is the number of uncorrelated constraints; $d_i$ and $\mathcal{Z}_i$ stand for the constraint and its SD; $\xi_i(\mathbf{\Theta})$ corresponds to the value the model defined by the parameter set $\mathbf{\Theta}$ provides for the quantity $i$. For the meaning and values of $d_i$ and $\mathcal{Z}_i$, see Table~\ref{tab:constraints}.

The log-likelihood function for correlated constraints reads 
\begin{align}
	\log \mathcal{L}_{q,\mathrm{corr.}} \propto -\chi_{q,\mathrm{corr.}}^2 =  
	-\frac{1}{2} \sum_{j=1}^{N_q^\mathrm{corr.}} \sum_{r=1}^{\mathfrak{N}_j} \sum_{s=1}^{\mathfrak{N}_j} \left(\mathrm{cov}^{-1}\right)_\mathit{rs}^{(j)} \delta \xi_r^{(j)} \delta \xi_s^{(j)},
	\label{eq:Chi2-Corr}
\end{align}
where $N_q^\mathrm{corr.}$ represents the number of correlated constraints; $\mathfrak{N}_j$ is the number of ``states'' among which the correlations are accounted for; $\delta {\xi}$ stands for the difference between the value of the constraint and the value provided by the model and
\begin{align}
    \begin{split}
        \mathrm{cov}_\mathit{rs} &= \frac{1}{P-1}\sum_{k=1}^P {\xi}_k(n_r) {\xi}_k(n_s)  \\
        &-\frac{1}{P-1} \sum_{k=1}^P {\cal \xi}_k(n_r)\, \frac{1}{P-1} \sum_{k=1}^P {\xi}_k(n_s),
    \end{split}
	\label{eq:Covariance}
\end{align}
is the covariance between the values of the quantity $\xi$ computed at densities $n_r$ and $n_s$. 
The index $k$ in Eq.~\eqref{eq:Covariance} runs over the number $P$ of individual calculations and we have taken into account the Bessel's correction since the mean and the covariance are determined from the same sample. 
For $\chi$EFT calculations in Ref.~\cite{Drischler_PRC_2021}, which we use here, $P=6$. 
As follows from Tables~\ref{tab:constraints} and \ref{tab:runs}, $\mathfrak{N}_j=3$ and $N_q^\mathrm{corr.} = 1$. 
The normalization factors in Eqs.~\eqref{eq:Chi2-NoCorr} and \eqref{eq:Chi2-Corr} are disregarded because they are not relevant for sampling from the posterior distributions (as long as they do not depend on the input parameters, which they do not in our case). 

The log-likelihood function for threshold constraints is
\begin{align}
    \log \mathcal{L}_{q,\mathrm{thresh.}} \propto -\chi_{q,\mathrm{thresh.}}^2 = 
    \begin{cases}
        0,         & \text{condition satisfied} \\
        -10^{10},  &\text{condition violated} 
     \end{cases}
\label{eq:Chi2-Thr}
\end{align}

Our primary method of Bayesian inference was affine invariant Markov Chain Monte Carlo (MCMC). 
For the details of the implementation, employed software and analysis of convergence, see Appendix~\ref{App:MCMC}. 
For a general discussion on how to treat constraints pertaining to physical quantities, see Section~3.2 of Ref.~\cite{Beznogov_PRC_2023}. 
For the motivation and details of the implementation of correlations, see Supplemental Materials of Ref.~\cite{Drischler_PRC_2021}. 

\subsection{Priors}
\label{ssec:Priors}

For the same technical reasons as in Paper~I, we employ ``mixed'' input parametrization. 
More precisely, instead of using as input parameters the 13 parameters of the effective interaction, we use three NEPs ($\sat{n}$, $\sat{E}$ and $\sym{J}$) and 8 parameters of the effective interaction ($D_3$, $\eff{C}$, $\eff{D}$, $t_4$, $t_5$, $\sigma$, $\beta$, $\gamma$), while the values of $x_4$ and $x_5$ are fixed to zero.
The remaining parameters of the effective interaction, $C_3$, $C_0$ and $D_0$, are computed according to
\onecolumngrid
\begin{align}
	&C_3=\frac{1}{\sigma {\sat{n}}^{\sigma+1}}\Biggl\{-\sat{E} + \frac{\hbar^2}{10m} \left( \frac{3 \pi^2 }{2}\right)^{2/3}\sat{n}^{2/3} 
	-\left( \frac{3 \pi^2}{2} \right)^{2/3} \sat{n}^{5/3} \left[
	\frac{2}{5} \eff{C} 
  	+\frac{3 t_4}{80}\left(3 \beta+2 \right) \sat{n}^{\beta}
  	+\frac{t_5}{40}\left( 3\gamma+2\right)\left( 5+4 x_5\right)\sat{n}^{\gamma}\right] \Biggl\},\nonumber\\
	&C_0=-\frac{1}{\sat{n}} \Biggl\{ 
  	\left( \frac{3 \pi^2}{2}\right)^{2/3} \frac{\hbar^2}{5 m}\sat{n}^{2/3}  +
  	\left(\sigma +1\right) C_3 \sat{n}^{\sigma+1} 
	+ \left( \frac{3 \pi^2}{2}\right)^{2/3} \sat{n}^{5/3} \left[ \eff{C}+\frac{9 t_4}{80} \left( \beta+\frac53\right) \sat{n}^{\beta} +\frac{3t_5}{80}\left( 5+4 x_5\right) \left( \gamma+\frac53\right) \sat{n}^{\gamma}\right]\Biggl\},
  	\label{eq:re-param}
\end{align}
\begin{align}
	&D_0=\frac{\sym{J}-J_k}{\sat{n}}-D_3 {\sat{n}}^{\sigma} +
	\left( \frac{3 \pi^2}{2} \right)^{2/3} {\sat{n}}^{2/3} \left[
   	-\left( \frac{\eff{C}}{3}+\eff{D}\right)
   	+ \frac18 t_4 x_4 \sat{n}^\beta - \frac{1}{24} t_5 \left( 5 x_5+4\right) \sat{n}^{\gamma}
  	\right],\nonumber
\end{align}
with $J_k=\hbar^2/6m \left(3\pi^2/2 \right)^{2/3}\sat{n}^{2/3}$. 
\twocolumngrid

As can be seen from Eqs.~\eqref{eq:CeffDeff_B}, fixing the values of $x_4$ and $x_5$ to zero does not affect the behavior of $\eff{\widetilde{C}}$ and $\eff{\widetilde{D}}$ provided that the prior range for $t_4$ and $t_5$ is wide enough. 
In contrast, some loss of generality occurs because $\sym{J}$ and $t_4$ become independent of each other, see Eqs.~\eqref{eq:Esym2} and \eqref{eq:tv4}.

The domain we consider for $\sigma$, $0.07 \leq \sigma \leq 1.1$, is narrower than the one considered in Paper~I, $0.01 \leq \sigma \leq 1.1$. 
While this limitation significantly simplifies the sampling, it has very limited impact on the posteriors, see Appendix~B of Paper~I. 

\renewcommand{\arraystretch}{1.05}
\setlength{\tabcolsep}{8.0pt}
\begin{table}
	\caption{Domains of the priors.}
	\centering
	\begin{tabular}{cccc}
		\toprule
		\toprule
		Parameter      & Units                          & Min.              & Max.             \\
		\midrule
		$\sat{n}$      & $\mathrm{fm^{-3}}$             & 0.14              & 0.18             \\
		$\sat{E}$      & MeV                            & $-16.9$           & $-14.9$          \\
		$\sym{J}$      & MeV                            & 22.8              & 38.8             \\
		$D_3$          & MeV $\mathrm{fm^{3+3\sigma}}$  & $-3\,000$         & $3\,000$         \\
		$\eff{C}$      & MeV $\mathrm{fm^{5}}$          & $-500$            & 2\,000           \\
		$\eff{D}$      & MeV $\mathrm{fm^{5}}$          & $-1\,000$         & 2\,000           \\
		$t_4$		   & MeV $\mathrm{fm^{5+3\beta}}$	& $-8\,000$			& 3\,000		   \\
		$t_5$		   & MeV $\mathrm{fm^{5+3\gamma}}$  & $-4\,000$		    & 4\,000		   \\
		$\beta$		   &  ---							& 0.07				& 1.1			   \\
		$\gamma$	   &  ---							& 0.07				& 1.1			   \\
		$\sigma$       &  ---                           & 0.07              & 1.1              \\
		\bottomrule
		\bottomrule
	\end{tabular}
	\label{tab:Prior}
\end{table}
\setlength{\tabcolsep}{2.0pt}
\renewcommand{\arraystretch}{1.0}

The domains of the priors are presented in Table~\ref{tab:Prior}. 
For the three NEPs, the input ranges are identical to those used in Paper~I; they cover 10 SDs symmetrically around the respective mean value, cf. Table~\ref{tab:constraints}. 
For $D_3$, $\eff{C}$, $\eff{D}$, $t_4$ and $t_5$ the ranges were chosen by the trial and error procedure to make the ranges as narrow as possible without (strongly) cutting the posterior. 
Unlike the situation of standard Skyrme, here the physical constraints on effective masses cannot be cast directly into the constraints on $\eff{C}$ and $\eff{D}$ as there are additional density-dependent terms there, see Eqs.~\eqref{eq:CeffDeff_B} and \eqref{eq:meff}. 
For $\beta$ and $\gamma$ we chose the same range as for $\sigma$. 

We have employed uniform priors in the ranges provided in Table~\ref{tab:Prior}. 

\section{Results}
\label{sec:Results}

Before discussing our results, we briefly mention the fitting quality of our runs. For all runs expect run~3, the maximum a posteriori value of $\chi^2$ is less than 1.0. For run~3 it is 3.2.
\subsection{Nuclear Matter}
\label{ssec:NM}

\begin{figure*}
	\centering
	\includegraphics[]{"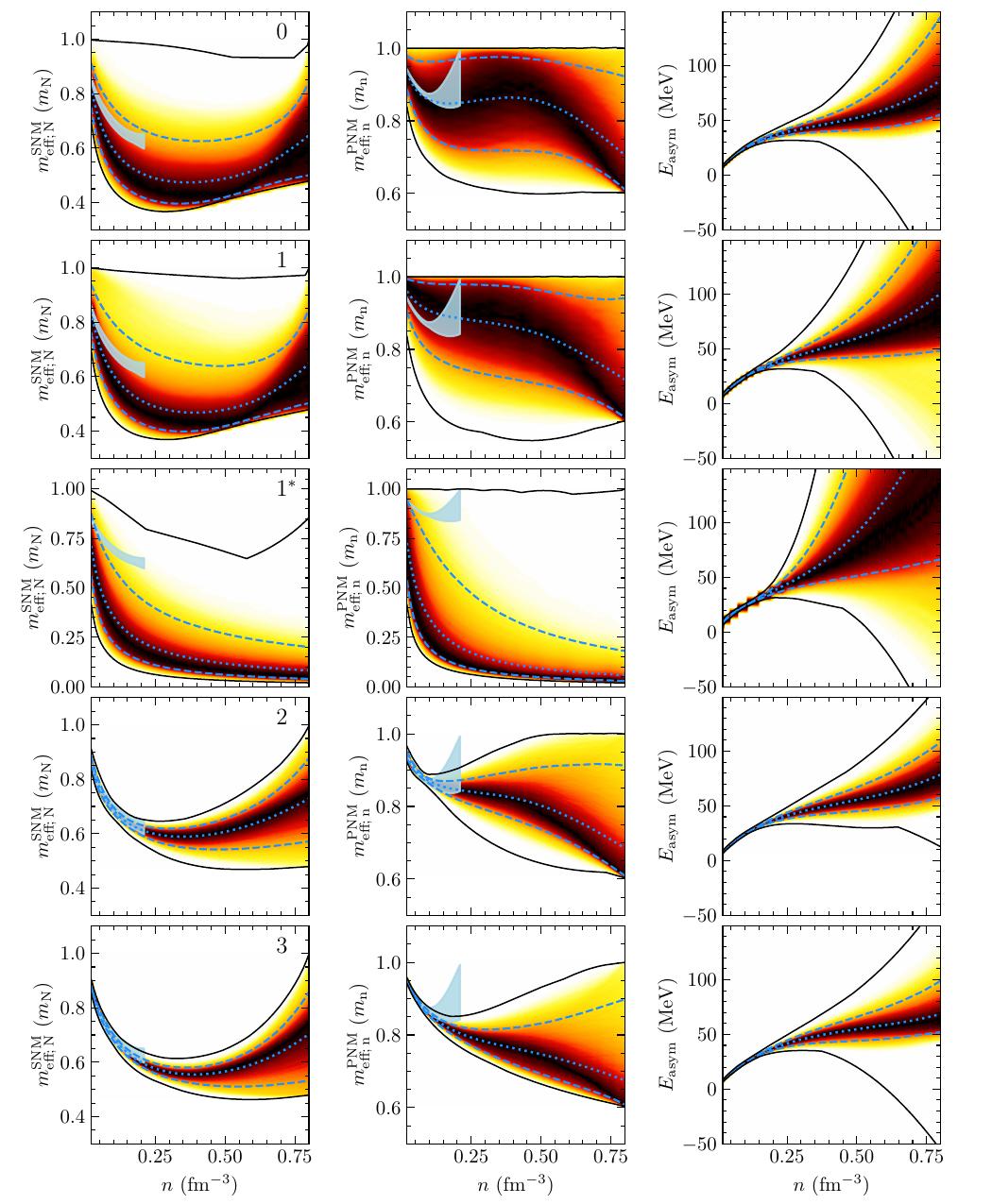"}
	\caption{Conditional probability density (also known as curve density) plots corresponding to the density dependence of the asymmetry energy, \eq{eq:Easym}, (right column), the nucleon effective mass in SNM (left column) and the neutron effective mass in PNM (middle column). 
    Results of various runs are illustrated on subsequent rows, as indicated in the left panels. 
    Curve density is indicated by colors: dark red (light yellow) corresponds to large (low) densities. 
    Dotted and dashed dodger blue curves demonstrate the median and 90\% CR, respectively. 
    Solid black curves mark the envelope of the bunch of models associated with each run. 
    Light blue shadowed regions mark the uncertainty domains, as computed in Ref.~\cite{Drischler_PRC_2021}.
	}
	\label{Fig:CD_NM}
\end{figure*}

The role that each constraint plays on NM is diagnosed following the modifications of conditional probability density corresponding to  the energy per nucleon in SNM and PNM; $\asym{E}$ [Eq.~\eqref{eq:Easym}]; $\mN$ and $\mn$ [Eq.~\eqref{eq:meff}], as functions of number density, as previously done in Paper~I. 
Though, at variance with Paper~I, here only the density dependence of $\asym{E}$; $\mN$; $\mn$ will be illustrated. 
The behavior of $E/A (n)$ in SNM and PNM will be nevertheless commented in detail.

Results pertaining to various runs in Table~\ref{tab:runs} are illustrated in Fig.~\ref{Fig:CD_NM} on subsequent rows. 
As in Paper~I, the conditional probability density plots of the posterior distributions are normalized such that for any given density the total probability is~1.0. 
The color map is normalized accordingly, with the black color corresponding to the maximum probability density at each given $n$.

The largest dispersion in $\left( E/A \right)_{\mathrm{SNM}}$ (not shown), $\left( E/A \right)_{\mathrm{PNM}}$ (not shown),
$\asym{E}$ (right), $\mN$ (left) and $\mn$ (middle column) are obtained for runs~1$^*$ and 1, which are the least constrained runs.

Accounting for correlations among $\left(E/A\right)_i$ in PNM (run~0) results in the suppression of models with extreme behaviors of $\left( E/A \right)_{\mathrm{PNM}}$ as a function of density, with more impact on the stiffest models than on the softest ones. 
As a stiff behavior of PNM relates to large values of $(\eff{\widetilde C}+ \eff{\widetilde D})$, by eliminating stiff PNM models, some of the stiff SNM models, characterized by large values of $\eff{\widetilde C}$, are also eliminated. 
Indeed, the 95\% quantile of $\left( E/A \right)_{\mathrm{SNM}}$ and the median in run~0 become less steep than in run~1; the 90\% confidence interval (CI) band gets narrower. 
As a result, the curves corresponding to $\asym{E}(n)$ get concentrated in a narrower band; the 95\% quantile and median also move to lower values. 

Upon imposing constraints on the Fermi velocity, Eq.~\eqref{eq:vF}, the models with strong increase of $E/A$ in both PNM and SNM as a function of density are disfavored.
The situation is understandable given that 
i) the upper limit on $v_\mathrm{F}$ translates into a lower limit on effective masses, and
ii) large values of $\left(E/A \right)_{\mathrm{PNM}}$ [$\left( E/A \right)_{\mathrm{SNM}}$] require large values of $(\eff{\widetilde C}+\eff{\widetilde D})$ [$\eff{\widetilde C}$] that, in turn, lead to low values of $\mn$ [$\mN$]. 
Soft models of $\left( E/A \right)_{\mathrm{SNM}}$ are disfavored as well. 
The impact of $v_{\mathrm{ F;\,n}}$ on $\asym{E}(n)$ is particularly strong, run~1 featuring a less pronounced increase with $n$ than run~1$^*$.

The comparison with the results produced by standard Skyrme interactions under similar constraints, i.e., run~1$^*$ here vs run~1 in Paper~I, allows one to see that
i) the increase of $\left( E/A \right)_{\mathrm{PNM}}$ with density is much stronger for extended Skyrme than for standard Skyrme,
ii) the increase of $\left( E/A \right)_{\mathrm{SNM}}$ with density is slightly weaker for extended Skyrme than for standard Skyrme.
In all present runs, models providing $\asym{E}$ increasing with density are dominant; this is in contrast with the results in Paper~I, where a significant fraction of the models in each run allowed $\asym{E}$ to become negative at high densities. 

Left and middle columns in Fig.~\ref{Fig:CD_NM}, where the behavior of $\mN$ and $\mn$ as functions of density is investigated, show that for run~1$^*$ the decrease of effective masses with density is much stronger than for standard Skyrme, see Fig.~2 in Paper~I, especially at low densities. 
This hints at a strong increase of $\eff{\widetilde C}(n)$ and $\eff{\widetilde D}(n)$.
The condition $v_{\mathrm {F;\,n}}/c \leq 1 $ prevents $\eff{m}$ from dropping too much.
The increased flexibility of the functional allows for a saturation in the decrease of effective masses as functions of density. 
The figure shows that, for densities in excess of $2\,\sat{n}$, $\mN$ increases with $n$; this behavior is in qualitative agreement with the predictions of microscopic models~\cite{Baldo_PRC_2014,Burgio_PRC_2020,Drischler_PRC_2021}.
The behavior of $\mn$ is more spectacular in the sense that, at least for run~0, many models manifest a second extremum, followed by a steep decrease.   
For runs~0 and 1, where only constraints on $\left(E/A\right)_i$ in PNM are implemented, the uncertainties bands are large for both $\mN$ and $\mn$.

Implementation of constraints on $\mN$ ($\mn$) suppresses models with either too large or too low values for this quantity. 
The panel corresponding to run~1 in Fig.~\ref{Fig:CD_NM} shows, for instance, that the number of models with large values of $\mN$ is smaller than the number of models with low values of $\mN$. 
These models correspond to cases with extreme values of $\eff{\widetilde C}$ ($\eff{\widetilde C}+ \eff{\widetilde D}$), i.e., to models with the softest or stiffest behavior of SNM (PNM). 
Upon their elimination, the envelopes of $\left( E/A \right)_{\mathrm{SNM}}(n)$ [$\left( E/A \right)_{\mathrm{PNM}}(n)$] and $\asym{E}(n)$ shrink considerably. 
Accounting additionally for correlations among the values that $\mni{i}$ takes at different densities (run~3) slightly favors models with stiffer increase of $E/A(n)$ in PNM and SNM, which results in an $\asym{E}$ envelope intermediate between those of run~0 and run~2.
For $\mN$, a good agreement with the data from Ref.~\cite{Drischler_PRC_2021}, which is used as constraints, is obtained up to $n=0.2~\mathrm{fm}^{-3}$. 
In what regards $\mn$, the situation is less good. 
For $n \lesssim 0.1~\mathrm{fm}^{-3}$, our data fit the constraints from Ref.~\cite{Drischler_PRC_2021} well, while at higher densities it undershoots the constraints. 
The explanation is straightforward: for $\mN$, the target $\chi$EFT band has roughly constant width, which means that all constraints contribute more or less equally. 
For $\mn$, the width of the target band is rapidly increasing with density, which results in low density ($\lesssim$0.08~fm$^{-3}$) constraints being significantly more important than the high density (0.16~fm$^{-3}$) ones. 
Then, our data also fail to provide $\mn$ that, for densities exceeding $\approx \sat{n}$, increases with density. 
Accounting for correlations among the values that $\mni{i}$ takes at different densities further shrinks the uncertainty bands for densities $n \lesssim 0.5$~fm$^{-3}$.
In the end, we mention that the run in which correlations among the values that $\mN$ takes at different densities are taken into account leads to a collection of models very similar to the one in run~2.

\subsection{Neutron stars matter}

\begin{figure*}
	\centering
	\includegraphics[]{"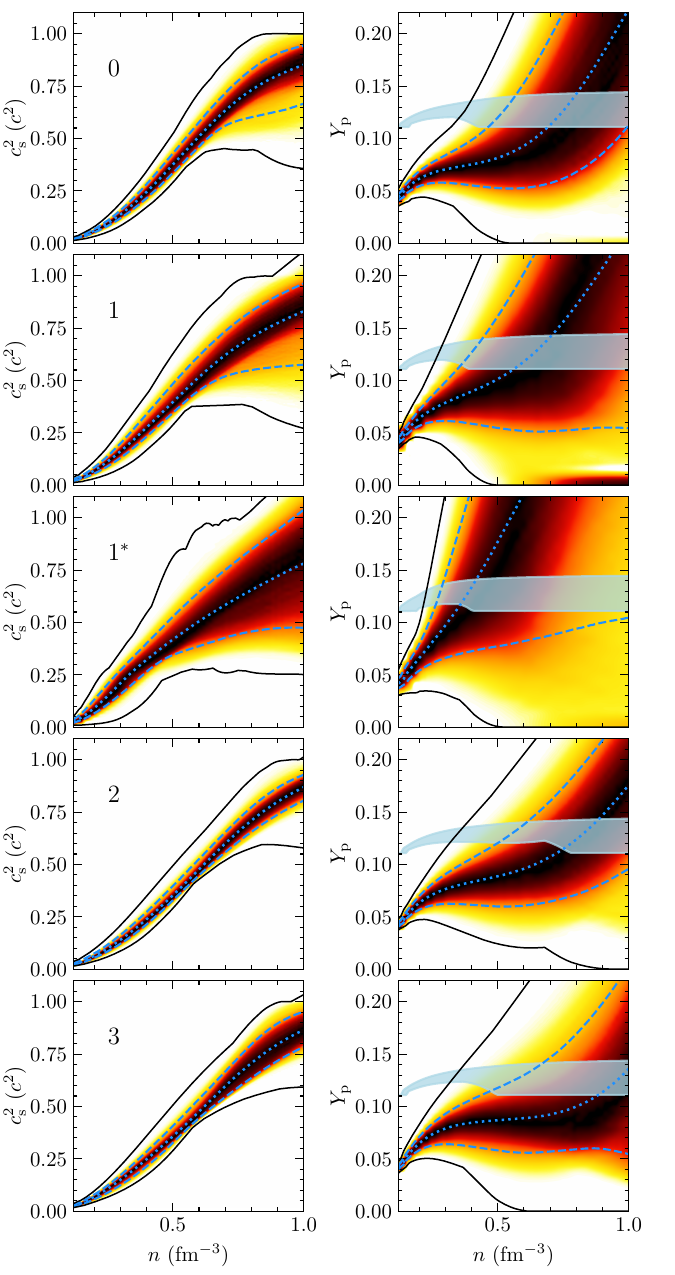"}
	\caption{Conditional probability density (also known as curve density) plots corresponding  to the density dependence of the speed of sound squared (left column) and proton fraction (right column) in NS matter. 
    Results of various runs are illustrated on subsequent rows, as indicated on the left panels. 
    The color map shows the curve density. 
    Dotted and dashed dodger blue curves demonstrate the median and 90\% CR, respectively.
    Solid black curves mark the envelope of the bunch of models associated with each run. 
    The light blue domains on the right panels correspond to the direct URCA threshold, as predicted by our models; see text for details.
	}
	\label{Fig:CD_NS}
\end{figure*}

We now turn to investigate the role that each set of constraints plays on the NS EOS. 
NS EOSs are obtained by smoothly matching the core and the crust EOSs. 
For the outer and inner crusts we adopt the models by Haensel, Zdunik and Dobaczewski~\cite{HDZ_1989} and Negele and Vautherin~\cite{NV_1973}, respectively. 
The matching between crust and core EOSs is done at a density of about $\sat{n}/2$. 
We assume that the leptonic sector consists of both electrons and muons.

Basic properties of NS EOSs are addressed in Fig.~\ref{Fig:CD_NS}. 
The left column depicts the density dependence of the speed of sound squared ($\cs^2$), defined as $\cs^2/c^2=\left(dP/de \right)_\mathrm{fr}$, where the subscript “fr” indicates that the derivatives have to be evaluated with the composition frozen. 
The composition of beta-equilibrated matter is considered in the right column in terms of the proton fraction $Y_{\mathrm p} = n_\mathrm{p}/n$. 
As before, each row corresponds to one of the runs in Table~\ref{tab:runs}. 

The models in the least constrained runs (run~1 and run~1$^*$) show, as expected, the largest dispersion.
The extra constraints on effective masses reduce the dispersion of $\cs^2(n)$-curves over the whole density range.
Accounting for correlations between the values that $E/A$ in PNM takes at different densities slightly reduces the dispersion.
The narrowest uncertainty band at 90\% CR corresponds to run~2; this is the situation where $\asym{E}$ curves show the smallest dispersion. 
Runs~0 and 1 contain models that soften at densities $n \sim 4 \sat{n}$, which makes that the lower boundary of the envelopes decreases with the density. 
The number of these models is nevertheless very small because the curves corresponding to the 5\% quantiles still increase as functions of density.
We also note that for densities exceeding a certain (model dependent) value, $\cs^2(n)$ tends to saturate. 
The strongest effect is seen for soft models in runs~0 and 1. 
None of the runs in Paper~I presented this behavior.
Comparison between results of runs~1 and 1$^*$ allows to judge the effect of the $v_{\mathrm{F;\,n}}$ constraint. 
It comes out that this condition suppresses the models with the most extreme values of $\cs^2$.

The $\asym{E}$ increase with $n$, shown by most of the models, see Fig.~\ref{Fig:CD_NM}, makes that for a huge majority of EOS models, $Y_{\mathrm p}$ of $\beta$-equilibrated matter increases with the density. As such, at variance with the results of Paper~I, here only a small fraction of models allow for NS cores made exclusively of neutrons. 
The uncertainty bands of all runs are still large, indicating that the isovector channel is not well constrained.
The shadowed light blue bands mark the threshold of the direct URCA processes,
$Y_{\mathrm {p;\,DU}}=1/\left[ 1+ \left(1+x_\mathrm{e}^{1/3} \right)^3 \right]$, with $x_\mathrm{e}=n_\mathrm{e}/\left( n_\mathrm{e}+n_{\mu} \right)$.
It comes out that all runs accommodate models that allow for this fast cooling process to operate as well as
models where this reaction is forbidden.
The constraint on $v_\mathrm{F;\,n}$ suppresses models with steep $Y_\mathrm{p}(n)$ behavior.

\subsection{Model dependence of the result}

Here, we investigate further the way in which the various sets of constraints modify the NM and NS EOSs. 
The analysis is done by considering 1D marginalized posterior distributions of different quantities. 
Examined are: parameters of NM; energy per particle in PNM with densities between 0.04 and 0.16~${\mathrm{fm}^{-3}}$; neutron (nucleon) effective mass in PNM (SNM) with $n = 0.16~{\mathrm{fm}^{-3}}$; global parameters of NSs.

\subsubsection{Posterior distributions of NM parameters; $E/A$ in PNM and effective masses}
\label{sss:NEPs}

Similarly to what was done in Paper~I and for the same reason of improving the efficiency of the sampling, the MCMC exploration is performed by controlling $\sat{n}$, $\sat{E}$ and $\sym{J}$, see Sec.~\ref{ssec:Priors}.
The first two quantities are allowed to take values in a narrow domain, see Table~\ref{tab:constraints}.
As such, their posterior distributions are confined and differ little from one run to the others. 
Posterior distributions of $\sat{E}$ sit on the top of the target distribution; with the exception of run~1$^*$, posterior distributions of $\sat{n}$ are slightly shifted toward higher values. 
The latter result suggests that the constraint imposed on $\sat{n}$ is in some tension with some of the constraints that govern the high density behavior, e.g., $\sat{K}$.

Posterior distributions of $\sat{X}^{i}$ with $i=2,3,4$ differ from one run to the others. 
The order of peaks' positions of $\sat{K}$ distributions is opposite to the one of $\sat{Z}$ distributions and similar to the one of $\sat{Q}$ distributions. 
This is due to a strong and negative correlation between $\sat{K}$ and $\sat{Z}$ and a loose and positive correlation between $\sat{K}$ and $\sat{Q}$, which was previously discussed in Refs.~\cite{Malik_ApJ_2022,Beznogov_PRC_2023,Malik_PRD_2023,Beznogov_ApJ_2024} and seems to be a universal feature. 
Run~3 provides the narrowest distributions of $\sat{K}$ and $\sat{Z}$. 
The narrowing of the $\sat{K}$ distribution upon accounting for correlations among $\mn$ is the out-turn of the coupling between the isovector and isoscalar sectors, see Fig.~\ref{Fig:Corner_run3}.
All the $\sat{K}$ distributions miss the target distribution.
The narrowest distributions of $\sat{Q}$ are obtained for runs~0 and 2. 
The relative positions of $\sat{K}$ and $\sat{Z}$ posteriors corresponding to runs~1, 0 and 2 suggest that accounting for correlations among the values that $E/A$ in PNM takes at different densities softens (stiffens) the EOS at densities slightly (a few times) higher than $\sat{n}$.  
$\sat{X}^i$ distributions are almost identical for runs~0 and 2, which means that the constraints imposed on the density dependence of $\mN$ and $\mn$ have on the isoscalar channel effects similar to those of accounting for correlations among $(E/A)_i$ in PNM. 
Accounting for correlations among the values that $\mn$ takes at different densities shifts $\sat{K}$ and $\sat{Q}$ ($\sat{Z}$) distributions to larger (lower) values. 
Posteriors of run~3 are peaked at same values as posteriors of run~1 but are narrower.

Posterior distributions of $\sym{J}$ are almost identical for runs~0 and 1 and runs~2 and 3. 
For runs~2 and 3, the peak value of $\sym{J}$ fits the peak value of the target distribution.
Posterior distributions of $\sym{X}^i$ with $i=1,...,4$ are different from one run to the others. 
The narrowest distributions of $\sym{L}$ corresponds to run~0; the narrowest distribution of $\sym{Z}$ corresponds to run~3.
These are the cases where the lowest (largest) peak values are obtained for $\sym{L}$ ($\sym{Z}$), too.
The fact that for $i=1,...,4$ $\sym{X}^i$ distributions of run~0 are narrower than for run~1 confirms that accounting for correlations among $E/A$ in PNM reduces the uncertainty band in the isovector channel. 
The effects of accounting for correlations among $\mn$ seem to be more important for high order terms in the Taylor expansion of the symmetry energy than for low order terms.

\begin{figure*}
	\centering
	\includegraphics[]{"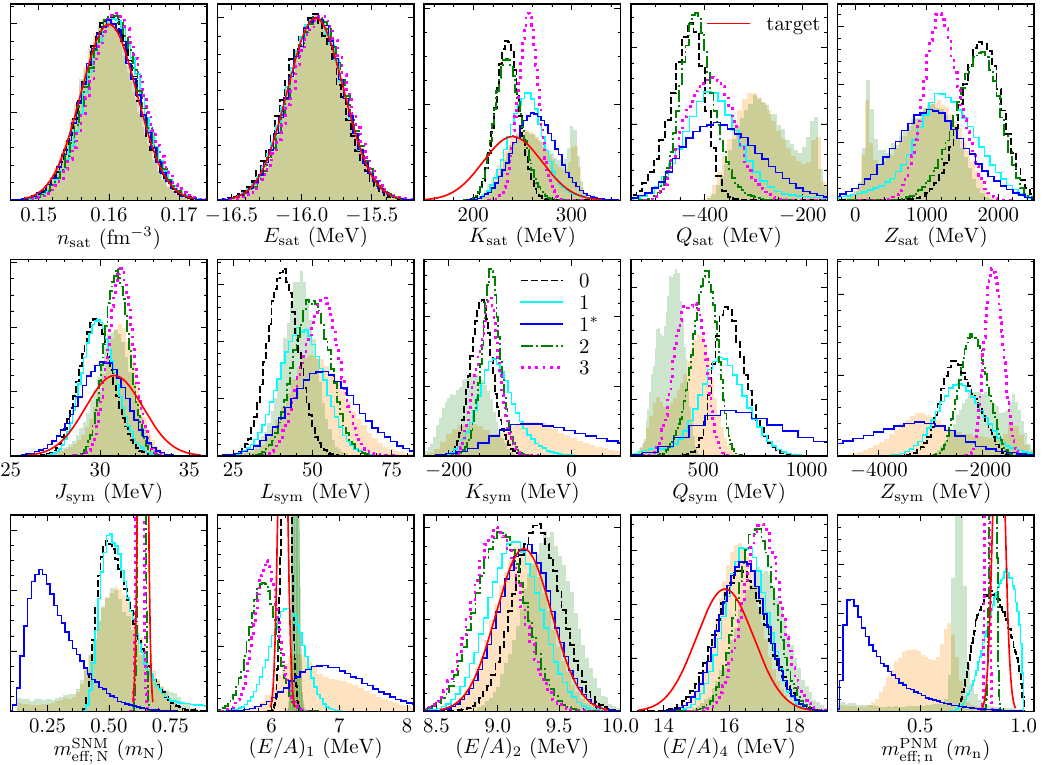"}
	\caption{Marginalized posteriors of NEPs; effective nucleon mass in SNM with density $n=0.16~{\mathrm{fm}}^{-3}$;
    effective neutron mass in PNM with density $n=0.16~{\mathrm{fm}}^{-3}$ and energy per particle in PNM with densities $n=0.04,~0.08$ and $0.16~{\mathrm{fm}}^{-3}$. 
    All runs are considered. 
    When available, target distributions for constrained parameters are also plotted. 
    For comparison, marginalized posteriors corresponding to runs~0 and 1 in Paper~I are illustrated as well (green and orange shaded areas, respectively). 
    Y-axis ranges have been chosen to increase readability. As such, some of the very narrow distributions are cut.
	}
	\label{Fig:Hist_NM}
\end{figure*}

The target distribution of $\mN$ at $n=0.16~\mathrm{fm}^{-3}$ is met only when the behavior of this quantity is explicitly constrained, i.e., in runs~2 and 3.
The target distribution of $\mn$ at $n=0.16~\mathrm{fm}^{-3}$ is never met, even if the posteriors of runs~2 and 3 are narrow and closer to the $\chi$EFT distribution than the posteriors of runs~0 and 1. 
As it was the case of $\mn(n)$ shown in Fig.~\ref{Fig:CD_NM}, this is the out-turn of the uncertainty band in $\chi$EFT calculations~\cite{Drischler_PRC_2021} that widens with density such that the constraints imposed at high densities are much less efficient than those imposed at low densities.
$\mN$ distributions of runs~0 and 1 are identical. 
This is a first indication that the isovector and isoscalar channels are less coupled than in Paper~I.
$\mn$ distributions are different for runs~0 and 1 and both are wide.
When compared with equivalent distributions corresponding to the standard Skyrme interactions, see Fig.~7 in Paper~I, one notes that here effective masses are prevented from taking small values.
This is the outcome of the constraint imposed on neutron Fermi velocity. 

$\chi$EFT distributions of $E/A$ in PNM with densities equal to 0.04, 0.08 and $0.16~\mathrm{fm}^{-3}$ are best described by run~0, which accounts for correlations among the values that this quantity takes at different densities. 
Even if no constraint is posed at $n=0.04~\mathrm{fm}^{-3}$, it is at this density that the predictions of run~0 agree best with those of $\chi$EFT~\cite{Drischler_PRC_2021}. 
We interpret this situation as the consequence of the uncertainty range of the $E/A$ constraints, which widens with the density. 
In addition, one can see that run~0 here matches $\chi$EFT results somewhat better than run~0 in Paper~I (green shaded area). The peak value of $(E/A)_1$ agrees with the peak value of the corresponding $\chi$EFT distribution also in run~1, though the posterior is much wider than the $\chi$EFT reference. 
Note that for run~1 in Paper~I (orange shaded area) the $(E/A)_1$ distribution misses the target distribution. 
The $(E/A)_4$-distributions of runs~1 here and in Paper~I are almost identical and narrower than the corresponding distribution of the more constrained run~0. 
Along with distributions of other runs, they span values higher than those of the $\chi$EFT distribution.  
Accounting for extra constraints on $\mn$ and $\mN$ makes all the $(E/A)_i$ distributions in Fig.~\ref{Fig:Hist_NM} deviate even more from the $\chi$EFT distributions. 
Together with the way in which the different constraints modify the distributions of $\mN$ and $\mn$, this suggests that the flexibility introduced by the extra terms in the Brussels functional is not enough to match the density behavior of both $E/A$ and $\eff{m}$ in $\chi$EFT calculations. 

All $\sat{X}^{i}$, $\sym{X}^i$, $\mni{i}$, $\mNi{i}$ and $(E/A)_i$ distributions are mono-modal. 
This situation is at variance with the one obtained in Paper~I, where many of these quantities, e.g., $\sat{K}$, $\sat{Q}$, $\sat{Z}$, $\sym{K}$, $\sym{Q}$, $\sym{Z}$, $(E/A)_1$, $\mn$, show bimodal distributions. 
In the case of the standard Skyrme interactions, the bimodality arises because the energy density functional becomes degenerate at $\sigma=2/3$. 
The extra momentum dependent terms in extended interactions lift the $\sigma$-related degeneracy and, thus, suppress the bimodality. 
Shaded areas in Fig.~\ref{Fig:Hist_NM}, that correspond to runs~0 and 1 in Paper~I, show that NEP distributions in this paper are, in most cases, narrower than those in Paper~I. 
Considering that the interactions used here have more parameters than the interactions in Paper~I, we attribute the better constraints of EOSs in this paper to the extra constraint on Fermi velocity. 
Further evidence in this sense is given by posteriors of run~1$^*$.  

When the constraint on the Fermi velocity~\cite{Urban_PRC_2023} is removed, a larger portion of the parameter space becomes available. 
This translates into a larger variety of models and, thus, broader distributions of NEPs. 
The modification of the $\sat{K}$, $\sat{Q}$, and $\sat{Z}$ posteriors with respect to run~1 is rather limited, which means that the isoscalar channel is not strongly limited by $v_\mathrm{F}$ constraints. 
The modification of $\sym{X}^{i}$ with $i=0,...,4$ are more important and increase with the order in the expansion. 
For the first time, the target distribution of $\sym{J}$ is described satisfactorily but this might be a coincidence. 
The fair agreement between $\sym{L}$ distributions in this run and run~1 in Paper~I might be fortuitous, too. 
The fact that the $\sym{Q}$ and $\sym{Z}$ distributions are wider than those of run~1 in Paper~I is the consequence of a larger number of parameters in the effective interaction parametrization. 
The fair agreement between the $(E/A)_1$ distributions in run~1$^*$ here and run~1 in Paper~I could be the out-turn of the fair agreement of $\sym{L}$ and $\sym{K}$ distributions. 
We remind that none of these runs posed constraints on the distribution of $(E/A)_1$. 
For the first time, the $(E/A)_2$ distribution perfectly fits the target distribution. 
The same holds for run~1 in Paper~I. 
The $(E/A)_4$ distribution resembles the distributions in runs~0 and 1. 
This result is in line with those corresponding to $\sat{X}^{i}$. 
Distributions of $\mn$ and $\mN$ are wide and peaked at very low values, i.e., $\sim 0.2 m_\mathrm{N}$; this are the situations where the largest discrepancies are obtained with respect to the other runs here and in Paper~I. 

In the end, we mention that the posteriors of a run similar to run~1 but with a larger value for the lower bound of the maximum gravitational mass ($M_{\mathrm{G}}^* \geq 2.2 \Msun$) (not shown) indicate a slightly stiffer behavior in both isovector and isoscalar channels. More precisely, posteriors of $\sat{K}$, $\sat{Q}$ and $\sym{K}$ become narrower and are centered at slightly higher values; the posterior of $\sat{Z}$ is narrower and shifted at slightly lower values; other distributions are marginally affected. This behavior is similar to the one shown by RMF models, see Fig.~6 in Ref.~\cite{Beznogov_PRC_2023}.

Medians and 68\% CI of marginalized posterior distributions of key NM quantities, including those plotted in Fig.~\ref{Fig:Hist_NM}, are provided in Table~\ref{tab:Posteriors} in Appendix~\ref{App:TableNMandNS}.

\subsubsection{Posterior distributions of global NS parameters}
\label{sss:NS}

%
\begin{figure*}
	\centering
	\includegraphics[]{"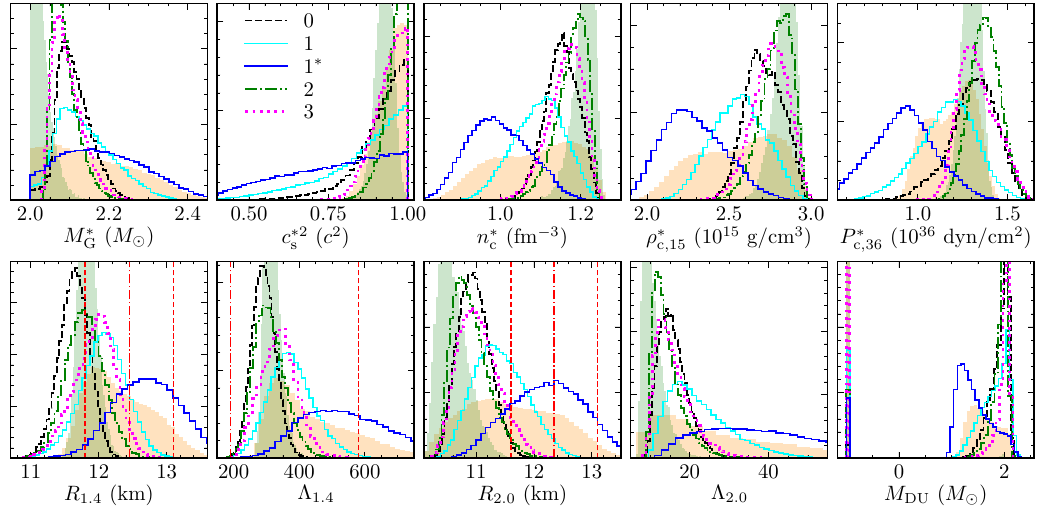"}
	\caption{Marginalized posteriors of selected properties of NSs. 
    Considered are: maximum gravitational mass ($M_{\rm G}^*$); the central density corresponding to the most massive configuration ($n_{\mathrm c}^*$); speed of sound squared ($c_{\rm s}^{*2}$), energy density ($\rho_{\rm c}^*$) and pressure ($P_{\rm c}^*$) at $n_{\mathrm c}^*$; radii ($R_{1.4}$, $R_{2.0}$) and tidal deformabilities ($\Lambda_{1.4}$, $\Lambda_{2.0}$) of NSs with masses equal to $1.4~\Msun$ and $2.0~\Msun$; the lowest mass of a NS that accommodates for direct URCA ($M_{\rm DU}$). 
    The vertical dotted-dashed and dashed lines on the $R_{1.4}$ and $R_{2.0}$ plots illustrate the median and the lower and upper bounds (at 68\% confidence), respectively, of the $R=12.45 \pm 0.65~\mathrm{km}$ and $R=12.35 \pm 0.75~\mathrm{km}$ constraints from Ref.~\cite{Miller_ApJL_2021}.
    The vertical dotted-dashed and dashed lines on the $\Lambda_{1.4}$ plot illustrate the median and the upper bound (at 90\% confidence), respectively, of the $190^{+390}_{-120}$ constraint from Ref.~\cite{Abbott_PRL_121}.
    For $M_{\rm DU}$ the value ``$-1$'' corresponds to the models that do not allow this process to operate in stable NSs. 
    For comparison, marginalized posteriors corresponding to runs~0 and 1 of Paper~I are illustrated as well (green and orange shaded areas, respectively). Y-axis ranges have been chosen to increase readability.
	}
	\label{Fig:Hist_NS}
\end{figure*}

Equilibrium configurations of static and spherically symmetric NSs are obtained by solving the Tolman-Oppenheimer-Volkoff equations. Tidal deformabilities are computed following Refs.~\cite{Hinderer_ApJ_2008,Hinderer_PRD_2010}.

Posterior distributions of a number of commonly considered global parameters of NS are illustrated in Fig.~\ref{Fig:Hist_NS}.
Posterior distributions for runs~0 and 1 in Paper~I are depicted for comparison. 
The 68\% CI combined constraints~\cite{Miller_ApJL_2021} on the radii of $1.4~\Msun$ and $2.0~\Msun$ NSs  are also shown on the corresponding panels. Constraints from GW170817 at 90\% CI, as obtained in Ref.~\cite{Abbott_PRL_121}, are illustrated on the plot corresponding to the tidal deformability of a $1.4~\Msun$ NS.

Posterior distributions of the maximum gravitational mass ($M_{\rm G}^*$) are asymmetric and peaked at values in excess of $2.0~\Msun$. 
In contrast, $M_{\rm G}^*$ distributions corresponding to runs~0 and 1 in Paper~I peak at $2.0~\Msun$, i.e., at the lower limit of the allowed maximum NS mass. 
Posterior distributions of the speed of sound squared at a density equal to the density of the maximum mass configuration ($c_{\mathrm s}^{*2}$) lean on the maximum accepted value, $c^2$. 
The $c_{\mathrm s}^{*2}$ distribution corresponding to run~0 of Paper~I peaks at $\sim 0.9 c^2$, while the one corresponding to run~1 is intermediate between distribution of run~0 of Paper~I and those obtained here.
When the constraint on $v_{\mathrm{F;\,n}}$ is accounted for, the widest (narrowest) distributions of $M_{\rm G}^*$ and $c_{\mathrm s}^{*2}$ correspond to run~1 (2). 

The explanation for the ``dip" in the $M_\mathrm{G}^*$ distribution, which ``shifts" the position of the peak, is that a significant number of models with the maximum mass close to the threshold value did not pass the criterion of having $c_{\mathrm s}^{*2}/c^2 \leq 1$ and, thus, have been suppressed. 
$M_\mathrm{G}^*$ distributions of runs~0 and 1 in Paper~I manifest no ``dip" and the respective $c_{\mathrm s}^{*2}/c^2$ posteriors are peaked at values lower than $c^2$. 
The posteriors of other quantities considered in Fig.~\ref{Fig:Hist_NS}, e.g., $n_{\mathrm c}^*$, $\rho_{\mathrm c}^*$, $R_{1.4}$, $R_{2.0}$, $\Lambda_{1.4}$, help to understand the reason of the discrepancy between present results and those in Paper~I. 
Posteriors of $n_{\mathrm c}^*$ and $\rho_{\mathrm c}^*$ for run~0 in Paper~I are concentrated in ranges that correspond to the highest values covered by the posteriors of the runs built here, which suggests that the EOSs in the run~0 in Paper~I are softer than most of the EOS models we built here. 
This conclusion agrees with what the values of $M_\mathrm{G}^*$ in run~0 in Paper~I, which are only slightly larger than $2~\Msun$, and a $c_{\mathrm s}^{*2}/c^2$ distribution peaked at $0.9-0.95$ convey. 
Posteriors of $n_{\mathrm c}^*$ and $\rho_{\mathrm c}^*$ for run~1 in Paper~I are broad and bimodal, which suggests that there a bunch of soft models coexists with a bunch of stiff models. 
The bunch of soft (stiff) models leads to $M_\mathrm{G}^*$ values slightly (significantly) larger than the threshold value; low (large) values of $R_{1.4}$, $R_{2.0}$, $\Lambda_{1.4}$ and $\Lambda_{2.0}$. 

The EOS models in runs~0, 2 and 3 have a tendency to be softer (stiffer) than most of the models corresponding to run~1 (0) in Paper~I; the models in runs~1 and 1$^*$ are stiffer than those in run~0 in Paper~I. 
The predictions of runs~0, 2 and 3 are relatively similar. Run~1, which is the least constrained, shows a wider range of behaviors.
EOSs' softness leads to $R_{1.4}$ and $R_{2.0}$ distributions that sit partially outside of the domains of Ref.~\cite{Miller_ApJL_2021}, which have been extracted based on a combination of pulsar timing data, X-ray data from NICER and XMM-Newton, GW data on tidal deformabilities and three theoretical frameworks of schematic EOS modeling. 
The disagreement with the data in Ref.~\cite{Miller_ApJL_2021} is more important for $R_{2.0}$ than for $R_{1.4}$. 
At variance with this, posteriors of $\Lambda_{1.4}$ for all runs except run~1$^*$ cover a domain delimited by the median and the upper boundary (at 90\% CI) from Ref.~\cite{Abbott_PRL_121}. 

When the very stringent constraint of the Fermi velocity~\cite{Urban_PRC_2023} is relaxed, the characteristics of EOS models change drastically. 
All posteriors become wide. 
The EOS models become stiffer than when this constraint is imposed, which explains why many models provide for $M_\mathrm{G}^*$ values that exceed the threshold value by much.
EOSs' stiffness also implies that distributions of $n_{\mathrm c}^*$, $\rho_{\mathrm c}^*$ and $P_{\mathrm c}^*$ span values lower than those in the other runs. 
The positions of the peaks of the $n_{\mathrm c}^*$ and $\rho_{\mathrm c}^*$ posteriors roughly coincide with the positions of the low-density mode of the run~1 in Paper~I. 
The fact that EOS models in run~1$^*$ are nevertheless different than those that make the ``stiff bunch" of run~1 in Paper~I is indicated by the $P_{\mathrm c}^*$ posterior, which does extend far beyond the range covered by the ``stiff bunch" of run~1 in Paper~I. 
In what regards $R_{1.4}$ and $R_{2.0}$, we note that the peaks are situated close to the median values from Ref.~\cite{Miller_ApJL_2021}. 
The peak of $\Lambda_{1.4}$ distribution is located somewhat below the 90\% upper bound from Ref.~\cite{Abbott_PRL_121} with $\approx 40\%$ of the posterior lying beyond this upper bound. 

Fig.~\ref{Fig:Hist_NS} also shows that all runs accommodate models that allow the direct URCA process to operate along with models where these reactions are forbidden.  
The percentages of models where the direct URCA acts in stable stars [stars with mass larger than $1.8~\Msun$] are: 98.7 [75.9], 88.9 [51.9], 93.8 [15.1], 93.2 [79.4] and 69.6 [64.9] for run~0, 1, 1$^*$, 2 and 3, respectively.

While there is no doubt that each constraint leaves imprints on every global parameter of NSs, the bunch of runs in this paper shows that the constraint on the Fermi velocity~\cite{Urban_PRC_2023} is by far the strongest constraint. 

Medians and 68\% CI of marginalized posterior distributions of key NS quantities, including those plotted in Fig.~\ref{Fig:Hist_NS}, are provided in Table~\ref{tab:Posteriors} in Appendix~\ref{App:TableNMandNS}.

The posteriors of the run similar to run~1 but with a larger value for the lower bound of the maximum gravitational mass ($M_{\mathrm{G}}^* \geq 2.2 \Msun$) (not shown) confirm that NSs' EOSs get stiffer. 
This behavior is consistent with the behavior of RMF models, see Fig.~6 in Ref.~\cite{Beznogov_PRC_2023}.

In the end, we mention that for all the ``output'' quantities plotted on Figs.~1 to 4 the posteriors of the run analogous to run~3 but with correlations between $\mNi{i}$ instead of $\mni{i}$ are very close to the posteriors of run~2. 
However, the posteriors of the parameters of the effective interactions (not included here) are different from those of run~2.
We explain the lack of effect of accounting for correlations among $\mNi{i}$ by the good constraint posed by individual values of $\mNi{i}$, in its turn, due to both narrow uncertainty band of $\chi$EFT data~\cite{Drischler_PRC_2021} and nearly constant width. The fact that different posteriors of the parameters of the effective interactions lead to almost identical ``output'' quantities illustrates inherent degeneracies of the energy density functional.

\subsubsection{Correlations between NEPs, $\mn$ and $\mN$}
\label{sssec:CorrelNM}

\begin{figure*}
  \centering
  \includegraphics[]{"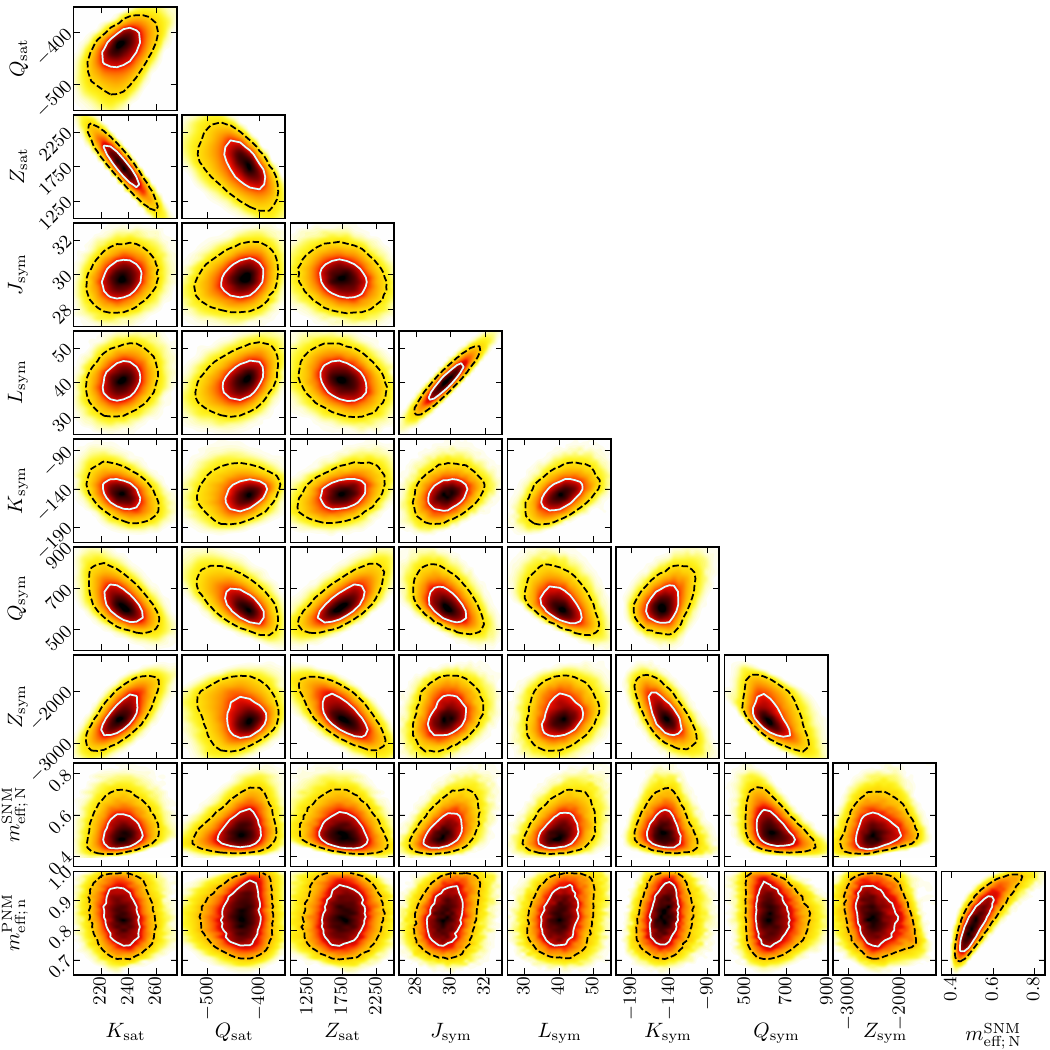"}
  \caption{2D marginalized posteriors of some of the NM parameters. The color map indicates the probability density. The light cyan solid and the black dashed contours show 50\% and 90\% CR, respectively. The results correspond to run~0 in Table~\ref{tab:runs}.}
  \label{Fig:Corner_run0}
\end{figure*}

\begin{figure*}
  \centering
  \includegraphics[]{"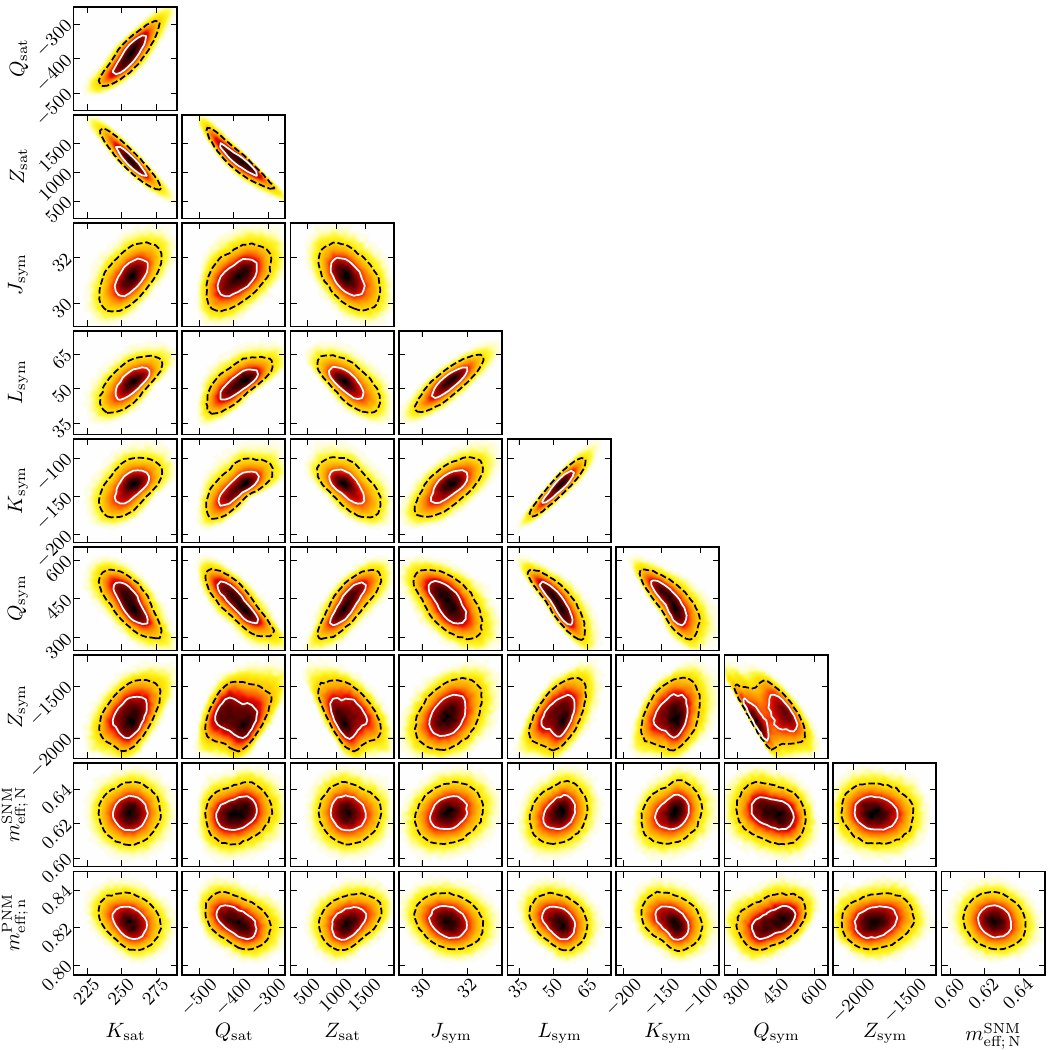"}
  \caption{The same as in Fig.~\ref{Fig:Corner_run0} but for run~3 in Table~\ref{tab:runs}.}
  \label{Fig:Corner_run3}
\end{figure*}

Correlations between NEPs, $\mn$ and $\mN$ corresponding to runs~0 and 3 in Table~\ref{tab:runs}
are illustrated in Figs.~\ref{Fig:Corner_run0} and \ref{Fig:Corner_run3}, respectively.

The only relatively strong correlations present in Fig.~\ref{Fig:Corner_run0} are $\sat{K}-\sat{Z}$ (negative), $\sym{J}-\sym{L}$ (positive) and $\mn-\mN$ (positive).
When compared with Fig.~9 in Paper~I that corresponds to run~0 in that paper, which is the equivalent of run~0 here, the situation is surprising.
None of the strong cross channel correlations present in Fig.~9 in Paper~I, e.g., $\sat{X}-\sym{Y}$ with $X,Y=K,Z,Q$, manifest here.
Then, the $\sat{Q}-\sat{Z}$ and $\sym{Q}-\sym{Z}$ correlations, which were very strong in Fig.~9 in Paper~I, barely manifest here. 
All the correlations in Paper~I that are lost when extra density dependent terms are included into the functional of the effective interaction are the spurious effect of the limited flexibility allowed by the standard Skyrme interaction.

When, instead of accounting for correlations between the values that $E/A$ in PNM takes at different densities, one accounts for the correlations between the values that $\mn$ takes at different densities, the correlation ``picture" sits in between the one showed by run~0 here and the one showed by run~0 in Paper~I.
Indeed, the correlations present in Fig.~\ref{Fig:Corner_run3} are:
$\sat{X}-\sat{Y}$ with $X,Y=K,Q,Z$, 
$\sym{X}-\sym{Y}$ with $X,Y=L,K,Q$, 
$\sat{X}-\sym{Y}$ with $X=K,Q,Z$ and $Y=L,K,Q$.
We also notice that the correlation $\mn-\mN$ present in Fig.~\ref{Fig:Corner_run0} does not manifest in Fig.~\ref{Fig:Corner_run3} and that the sign of the $\sym{K}-\sym{Q}$ correlation in run~3 here is opposite to the one in run~0 here and in Paper~I.

The correlation patterns of the run analogous to run~3 but with correlations between $\mNi{i}$ instead of $\mni{i}$ (not shown)
are qualitatively similar to those in Fig.~\ref{Fig:Corner_run3}.

The obvious conclusion of this investigation is that correlations among parameters of NM are very sensitive to the structure of the functional and the way in which the constraints are imposed.


\subsubsection{Correlations between parameters of NM and NS}
\label{sssec:CorrelNM-NS}

\begin{figure*}
  \centering
  \includegraphics[]{"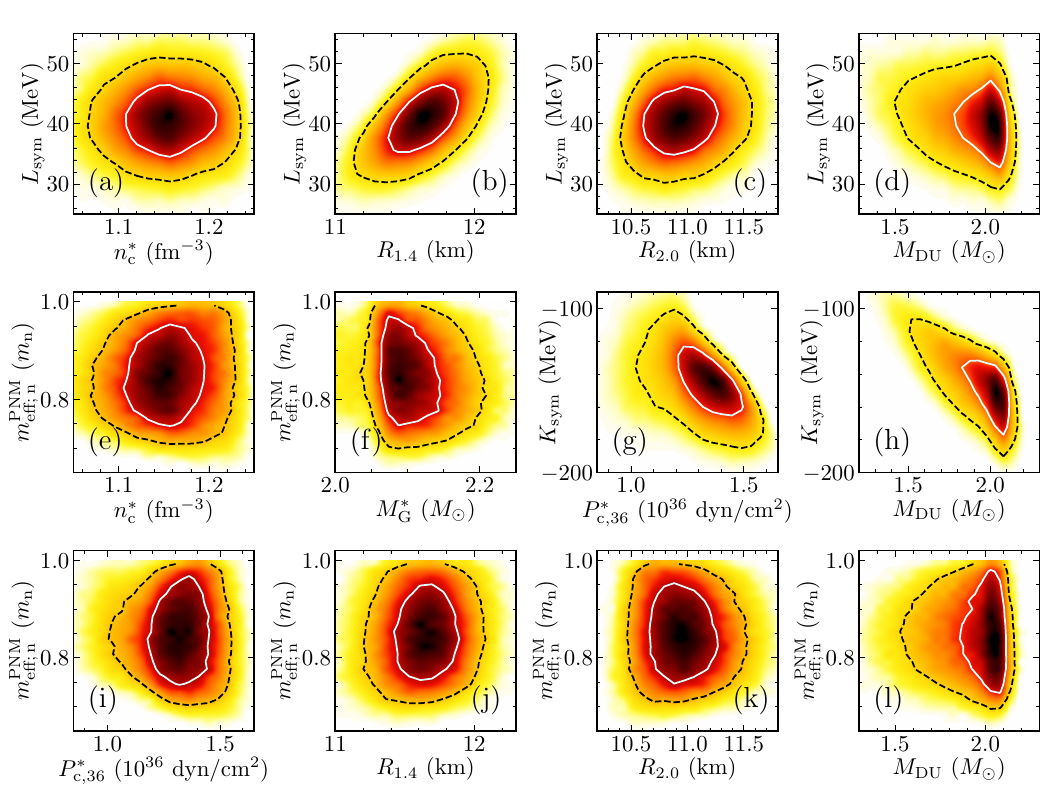"}
  \caption{2D marginalized posterior distributions for selected NM and NS parameters for which correlations have been previously obtained for run~1 in Ref.~\cite{Beznogov_ApJ_2024}. The color map indicates the probability density. The light cyan solid and the black dashed contours demonstrate 50\% and 90\% CR, respectively. Results correspond to run~0 in Table~\ref{tab:runs}.}
  \label{Fig:Correl_NS-NM_run0}
\end{figure*}

\begin{figure*}
  \centering
  \includegraphics[]{"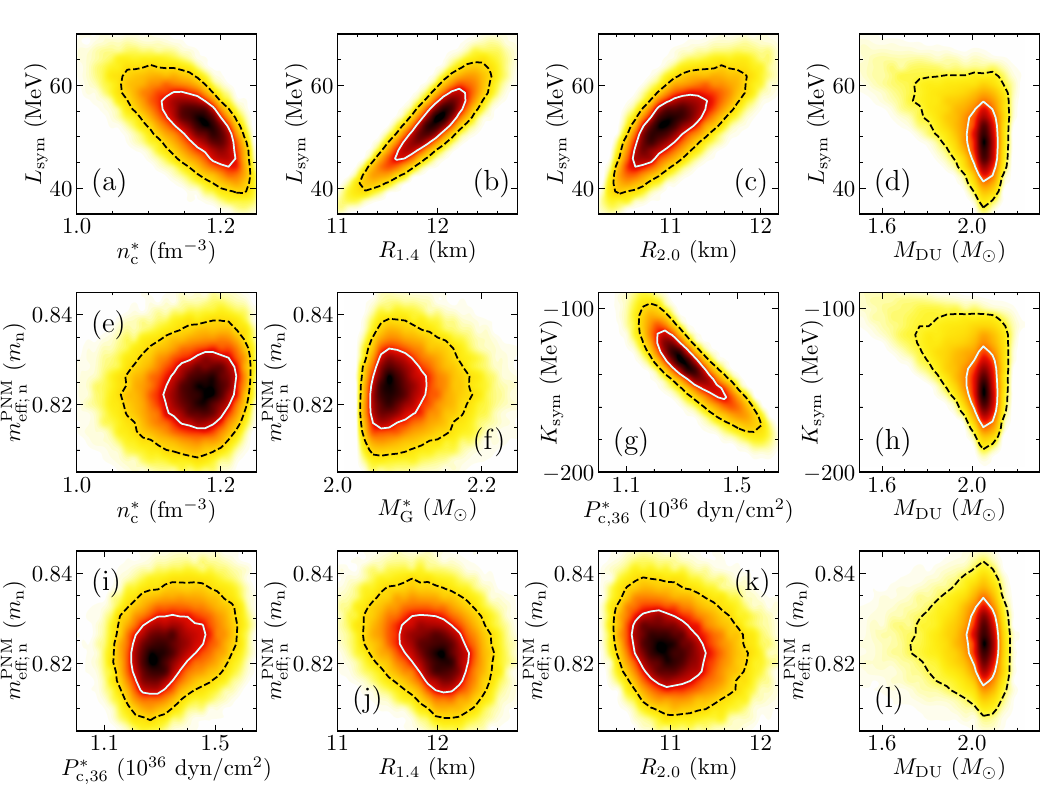"}
  \caption{The same as in Fig.~\ref{Fig:Correl_NS-NM_run0} but for run~3 in Table~\ref{tab:runs}.}
  \label{Fig:Correl_NS-NM_run3}
\end{figure*}

Correlations between selected NM and NS parameters corresponding to runs~0 and 3 in Table~\ref{tab:runs} are illustrated in Figs.~\ref{Fig:Correl_NS-NM_run0} and \ref{Fig:Correl_NS-NM_run3}, respectively. Considered are the combinations of quantities for which correlations have previously been obtained in Paper~I.

Similarly to what we have noticed with respect to Figs.~\ref{Fig:Corner_run0} and \ref{Fig:Corner_run3}, the augmented flexibility of extended Skyrme interactions washes out many of the correlations. The effect is much stronger for run~0 than for run~3.
Only the correlations $\sym{L}-R_{1.4}$, $\sym{K}-P_{\mathrm c}^*$ and $\sym{K}-M_{\mathrm{DU}}$ manifest in Fig.~\ref{Fig:Corner_run0} and all of them are weaker than those that manifest in Fig.~11 of Paper~I.
When, in exchange of accounting for correlations between $(E/A)_i$ in PNM, one accounts for correlations between $\mni{i}$ in PNM, the correlations $\sym{L}-R_{1.4}$ and $\sym{K}-P_{\mathrm c}^*$ get stronger, the correlation $\sym{K}-M_{\mathrm{DU}}$ disappears and a loose correlation manifest between $\sym{L}$ and $R_{2.0}$.

It comes out that, as it was the case with the correlations among parameters of NM, the correlations between parameters of NM and NSs are very sensitive to the structure of the functional and the way in which the constraints are imposed.
It is, thus, clear that in order for any correlation to be physical it has to manifest in any model in which the considered quantities are allowed to explore ranges that are wide enough.


\section{Conclusions}
\label{sec:Concl}

In this work, we have performed a full Bayesian investigation of the dense matter EOS built upon the non-relativistic mean field model of nuclear matter with Brussels extended Skyrme interactions.
Knowledge from nuclear structure experiments was implemented via constraints on the values of the four best known NEPs, i.e., $\sat{n}$, $\sat{E}$, $\sat{K}$ and $\sym{J}$. 
Knowledge from ab initio theoretical calculations of nuclear matter was incorporated via the density behavior of $E/A$ and $\mn$ in PNM and the density behavior of $\mN$ in SNM, respectively, as computed in Ref.~\cite{Drischler_PRC_2021} based on $\chi$EFT with NN interactions computed at N$^3$LO and 3N interactions computed at N$^2$LO.
The behavior of neutron rich matter with densities in excess of several times $\sat{n}$ was controlled by requiring that all EOS models comply with the $2~\Msun$ constraint on the lower bound of the maximum NS gravitational mass. NS EOSs were also required to be thermodynamically stable, i.e., $P >0$ and $dP/dn \geq 0$, and causal up to the density corresponding to the central density of the maximum mass configuration. 

In addition to these, we implemented a condition that is mandatory for Skyrme models to be physical, but brought into attention only recently by Duan and Urban~\cite{Urban_PRC_2023}. 
In consists in requiring that the Fermi velocity of nucleons in dense matter is inferior to the speed of light, which translates into a condition on the minimum value of the Landau effective masses. 
Our calculations show that this condition drastically changes the properties of NM and NSs.
The density dependence of the Landau effective masses acquires a behavior that, up to the densities of the order of $2-3 \sat{n}$, qualitatively reproduces the one predicted by ab initio calculations~\cite{Baldo_PRC_2014,Burgio_PRC_2020,Drischler_PRC_2021}. 
NS models show lower values of NS radii and tidal deformabilities and
larger values of $n_{\mathrm c}^*$, $\rho_{\mathrm c}^*$, $P_{\mathrm c}^*$. 
Moreover, extreme values of maximum NS gravitational masses ($\gtrsim 2.3~\Msun$) are disfavored.
We note that increasing the value of the density up to which the condition $v_{F;\,n}<c$ is enforced does not guarantee that the NS EOS is physical over the whole density range span in massive NS, which can be seen as a failure of non-relativistic models at densities in excess of $2\sat{n}-3\sat{n}$.    

Families of EOS models have been built for various sets of constraints. 
In addition to accounting for correlations among the values that $E/A$ in PNM takes at different densities, previously discussed in Paper~I, here we have also considered the modifications brought by accounting for the correlations among the values that $\mn$ ($\mN$) in PNM (SNM) takes at different densities. 
The behavior of NM under different sets of constraints was analyzed considering the behaviors of $E/A$ and effective masses as functions of density in PNM and SNM as well as the density dependence of $\asym{E}$. 
As is easy to foresee, any extra condition on $\mN$ and/or $\mn$ helps to narrow down the uncertainty band of the EOS, especially at low densities. 
NS EOSs were investigated considering the density dependence of the speed of sound squared and $Y_{\mathrm p}$.
The model dependence of the results; the role of individual constraints; the effects of extra momentum dependent terms were judged upon by confronting the posterior distributions of various NM and NS quantities.
For all the runs built here, the distributions of $n_{\mathrm c}^*$ extend up to densities exceeding 1.2~fm$^{-3}$, which are supposedly beyond the validity limit of the model. 
Strong core compression, typical to ``soft" models, is reflected also in relatively low values of $R_{1.4}$, $\Lambda_{1.4}$ and $R_{2.0}$. 
Similarly to the case of standard Skyrme, all of the runs here contain models that allow for direct URCA to operate in stable stars.

It is particularly interesting to notice that, under identical constraints, some of the predictions of Brussels extended interactions differ from those of standard Skyrme interactions~\cite{Beznogov_ApJ_2024}:
i) the overwhelming majority of Brussels interactions favor an (a)symmetry energy that increases with density, which translates into a $Y_{\mathrm p}$ of $\beta$-equilibrated matter also increasing with density;
ii) stiffer NS EOS;
iii) much lower values of $\mN$ and $\mn$ at $0.16~\mathrm{fm}^{-3}$;
iv) different distributions of $\sat{Q}$, $\sym{K}$, $\sym{Q}$, $\sym{Z}$. 
The last feature suggests that domains of values extracted by studying the behavior of either of these interactions should not a priori be imposed to the other. 

The augmented flexibility of the Brussels-Skyrme effective interactions with respect to the standard Skyrme ones drastically changes the patterns of correlations among NM parameters or among parameters of NM and NSs.
The strong cross channel correlations that appear in Paper~I upon accounting for correlations between $(E/A)_i$ in PNM do not manifest in the case of Brussels interactions.
In exchange, the isovector and isoscalar channels appear to get coupled when correlations between $\mni{i}$ are accounted for.
We also note that the sign of the $\sym{K}-\sym{Q}$ correlation for run~3 here is opposite to the one previously seen for run~0 in Paper~I.
The situation of correlations among parameters of NM and NSs is qualitatively similar.
With the exception of $\sym{L}-R_{1.4}$, $\sym{K}-P_{\mathrm c}^*$ and $M_{\mathrm{DU}}-\sym{K}$, none of the correlations present in run~0 in Paper~I manifest in run~0 here.
Switching from accounting for correlations between $(E/A)_i$ in PNM to accounting for correlations between $\mni{i}$ helps the correlations cited above to get stronger and new correlations, e.g., $\sym{L}-n_{\mathrm c}^*$, $\sym{L}-R_{2.0}$, to manifest.

\looseness=-1
The present paper is the follow-up of Paper~I; the two papers aim to explore the capabilities of the non-relativistic mean field model with Skyrme interactions to describe NS matter. 
Together with Refs.~\cite{Malik_ApJ_2022,Beznogov_PRC_2023,Malik_PRD_2023}, Paper~I and the present paper contribute to a wider investigation of the high density behavior of phenomenological mean-field models with nucleonic degrees of freedom. 
Contrary to other Bayesian studies in the literature that also use mean-field models, e.g., Refs.~\cite{Traversi_ApJ_2020,Char_PRD_2023,Papakonstantinou_PRC_2023,Imam_PRD_2024}, we have employed a minimum set of constraints and favored constraints from nuclear physics.
The obvious drawback of not implementing constraints from measurements of NSs' radii and tidal deformabilities resides in a loose constraint of the intermediate density domain of the EOS.

In Ref.~\cite{Beznogov_PRC_data_2024} we provide the posterior distributions of the input parameters for all the effective interactions that make our five main runs.

\begin{acknowledgments}
We express our gratitude to Micaela Oertel for drawing into our attention physicality issues related to the Fermi velocity of the nucleons exceeding the speed of light at densities in excess of $\sat{n}$, which plague many Skyrme interactions~\cite{Urban_PRC_2023}.

Support from a grant of the Ministry of Research, Innovation and Digitization, CNCS/CCCDI – UEFISCDI, Project No. PN-IV-P1-PCE-2023-0324 is acknowledged. Partial support from PN 23 21 01 02 is acknowledged as well.

M.V.B. and A.R.R. contributed equally to this work.
\end{acknowledgments}

\appendix
\section{Nuclear Empirical Parameters}
\label{App:NEPs}

We provide here analytic expressions for low order NEPs. 
A hybrid set of parameters mixing the $C_0$, $D_0$, $C_3$, $D_3$, $\eff{C}$, and $\eff{D}$ parameters, introduced in the standard Skyrme interactions in order to reduce the dimension of the parameter space relevant for the description of NM~\cite{Ducoin_NPA_2006}, with the $t_4$, $x_4$, $t_5$, and $x_5$ parameters of the extra terms along with $\sigma$, $\beta$, and $\gamma$ exponents that regulate the density dependence of the momentum dependent terms is employed. 
There are two arguments in favor of this choice. 
The first relates to the conciseness. 
More precisely, for each NEP, the expression corresponding to Brussels interactions $X^{\mathrm{BSk}}$ will be presented as a sum between a term entering the standard Skyrme interactions $X^{\mathrm{Sk}}$, provided in Paper~I, and extra terms specific to the extended interactions.
The second argument pertains to the way in which the parameter space exploration is done numerically, see Sec.~\ref{sec:Bayes}. 
For equivalent expressions in terms of the most commonly used $t_i$, $x_i$ with $i=0,...,4$ and $\sigma$, $\beta$, $\gamma$ coefficients, see Ref.~\cite{Dutra_PRC_2012}. 

The saturation energy of NM with isospin asymmetry $\delta$ writes: 
\begin{equation}
  \sat{E}^{\mathrm{BSk}} \left(\delta\right) = \sat{E}^{\mathrm{Sk}} \left(\delta\right) + {\cal T}_4 + {\cal T}_5 .
  \label{eq:Esat} 
\end{equation}

The corresponding expressions for incompressibility $\sat{K}\left(\delta\right)=\sat{X}^{\delta;\,2}$, skewness $\sat{Q}\left(\delta\right)=\sat{X}^{\delta;\,3}$ and kurtosis $\sat{Z}\left(\delta\right)=\sat{X}^{\delta;\,4}$ are:
\onecolumngrid
\begin{align}
  &\sat{K}^{\mathrm{BSk}} \left(\delta\right) = \sat{K}^{\mathrm{Sk}} \left(\delta\right)
  +  \left( 5+3 \beta \right) \left(8+3\beta \right)  {\cal T}_4
    + \left( 5+3 \gamma \right) \left(8+3 \gamma\right) {\cal T}_5 , 
  \label{eq:Ksat} \\
  &\sat{Q}^{\mathrm{BSk}} \left(\delta\right) = \sat{Q}^{\mathrm{Sk}} \left(\delta\right)
  + \left( 2+3 \beta \right)  \left( 5+3 \beta \right) \left(3\beta -1\right)  {\cal T}_4
    + \left( 2+3 \gamma \right) \left( 5+3 \gamma \right) \left(3 \gamma-1\right) {\cal T}_5 , 
 \label{eq:Qsat} \\
  &\sat{Z}^{\mathrm{BSk}} \left(\delta\right) = \sat{Z}^{\mathrm{Sk}} \left(\delta\right)
  + \left( 3 \beta -4\right) \left( 2+3 \beta \right)  \left( 5+3 \beta \right) \left(3\beta -1\right)  {\cal T}_4
    + \left( 3 \gamma -4\right) \left( 2+3 \gamma \right) \left( 5+3 \gamma \right) \left(3 \gamma-1\right) {\cal T}_5 .
 \label{eq:Zsat} 
\end{align}

In the above expressions we made use of $\tau_i(T=0)=\pi^{4/3} \left(3 n_i\right)^{5/3}/5$ and have introduced the notations:
\begin{align}
  &{\cal T}_4= \frac{3}{40} \left( \frac{3 \pi^2}{2}\right)^{2/3} \left(\sat{n}^{\delta}\right)^{5/3+\beta} 
  \left[ t_4 \left(x_4+2 \right)G_{5/3}-t_4 \left(x_4+\frac12 \right)G_{8/3}(\delta) \right],\\
  &{\cal T}_5=\frac{3}{40} \left( \frac{3 \pi^2}{2}\right)^{2/3} \left(\sat{n}^{\delta}\right)^{5/3+\gamma} \left[t_5 \left(x_5+2 \right)G_{5/3}+t_5 \left(x_5+\frac12 \right)G_{8/3}(\delta) \right],
\end{align}
and $G_\alpha(\delta)=\left[ \left(1-\delta \right)^\alpha+\left(1+\delta \right)^\alpha\right]/2$.

The symmetry energy can be computed as
\begin{align}
  \begin{split}
    E_{\mathrm{sym};\,2}^{\mathrm{BSk}} = E_{\mathrm{sym};\,2}^{\mathrm{Sk}}+ \mathfrak{T}_4
    +\mathfrak{T}_5,
  \end{split}
  \label{eq:Esym2}
\end{align}
while its slope $\sym{L}$, curvature $\sym{K}$, skewness $\sym{Q}$ and kurtosis $\sym{Z}$ write
\begin{align}
  &\sym{L}^{\mathrm{BSk}} = \sym{L}^{\mathrm{Sk}} + \left( 5+3\beta \right) \mathfrak{T}_4
  +\left( 5+3\gamma \right)\mathfrak{T}_5,
  \label{eq:Lsym} \\
  &\sym{K}^{\mathrm{BSk}} =\sym{K}^{\mathrm{Sk}} + \left( 5+3\beta \right) \left( 2+3\beta \right) \mathfrak{T}_4
  +\left( 5+3\gamma \right) \left( 2+3\gamma \right)  \mathfrak{T}_5,
  \label{eq:Ksym} \\
  &\sym{Q}^{\mathrm{BSk}} =\sym{Q}^{\mathrm{Sk}}+
  \left( 5+3\beta \right) \left( 2+3\beta \right) \left(3\beta -1 \right) \mathfrak{T}_4
  +\left( 5+3\gamma \right) \left( 2+3\gamma \right) \left(3\gamma -1 \right)  \mathfrak{T}_5,
  \label{eq:Qsym} \\
  &\sym{Z}^{\mathrm{BSk}} =\sym{Z}^{\mathrm{Sk}}+
   \left( 5+3\beta \right) \left( 2+3\beta \right) \left(3\beta -1 \right) \left(3\beta -4 \right) \mathfrak{T}_4
  +\left( 5+3\gamma \right) \left( 2+3\gamma \right) \left(3\gamma -1 \right)  \left(3\gamma -4 \right)   \mathfrak{T}_5 . 
  \label{eq:Zsym}
\end{align}
\twocolumngrid

In the above expressions we have introduced
\begin{align}
   &\mathfrak{T}_4=-\frac18 \left(\frac{3 \pi^2}{2}\right)^{2/3} t_4 x_4 n^{5/3+\beta},
   \label{eq:tv4}\\
   &\mathfrak{T}_5=\frac{1}{24} \left(\frac{3 \pi^2}{2}\right)^{2/3} t_5 \left( 5 x_5+4\right) n^{5/3+\gamma} 
   \label{eq:tv5}.
\end{align}

Notice that in Eqs.~\eqref{eq:Lsym} -- \eqref{eq:Zsym} the ``0" superscript has been dropped off for convenience. 
The same convention is used throughout this paper.

\section{MCMC implementation details}
\label{App:MCMC}

%
\begin{figure*}
	\centering
	\includegraphics[]{"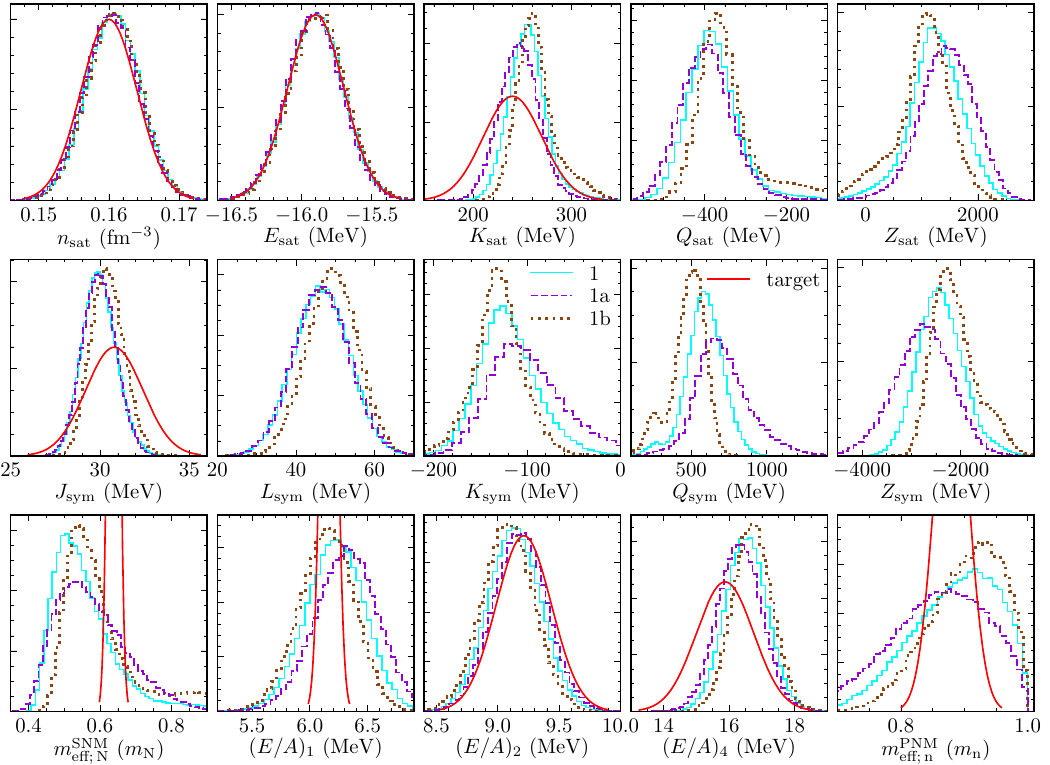"}
	\caption{The same as Fig.~\ref{Fig:Hist_NM} but for runs~1, 1a, and 1b. See Appendix~\ref{App:vF} for details.}
	\label{Fig:Hist_NM_vF}
\end{figure*}
\begin{figure*}
	\centering
	\includegraphics[]{"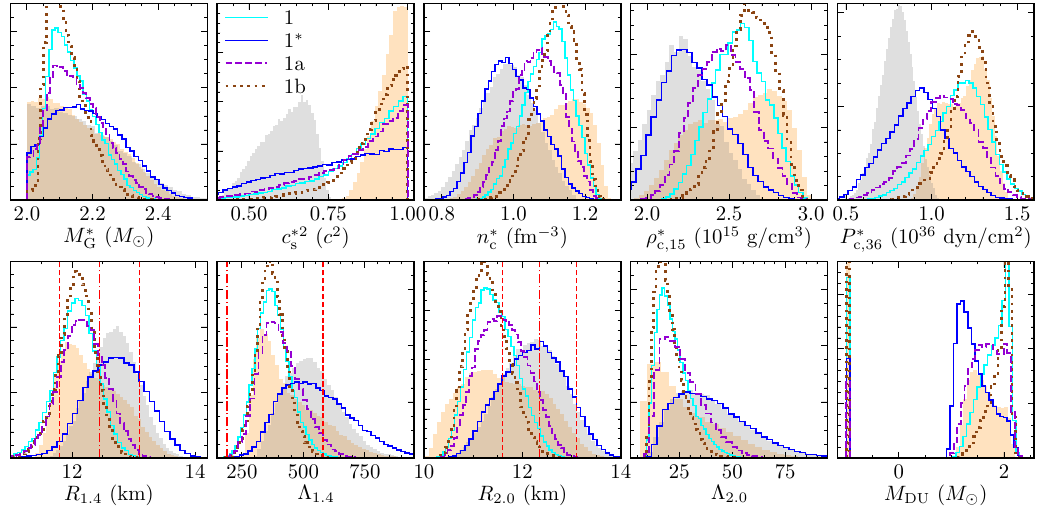"}
	\caption{The same as Fig.~\ref{Fig:Hist_NS} but for runs~1, 1$^*$, 1a, and 1b. Comparison with run~1 of Paper~I (orange shaded area) and with run~4 of Ref.~\cite{Beznogov_PRC_2023} (gray shaded area) are also presented. Y-axis ranges have been chosen to increase readability. See See Appendix~\ref{App:vF} for details.}
	\label{Fig:Hist_NS_vF}
\end{figure*}

The general approach to the MCMC setup and its implementation here follow closely those described in Appendix~B of Paper~I.
However, the higher dimensionality of the parameter space and the additional constraints on neutron Fermi velocity make the inference even more difficult.

We have employed \textsc{emcee} (v.3.1.4) Python package~\cite{emcee,emcee3}\,\footnote{The documentation is available at \url{https://emcee.readthedocs.io} and the GitHub page is \url{https://github.com/dfm/emcee}.} as an implementation of the affine invariant MCMC~\citep{Goodman_2010} and as our main Bayesian inference tool. 
In all calculations we have used kernel density estimate (KDE) steps and 1\,000 walkers. 
For the details of the two-stage inference procedure, autocorrelation length analysis and bootstrap analysis of the stability of the posteriors, see Appendix~B of Paper~I. 
Here, we focus on what is different compared to Paper~I.

First, due to the much higher computational cost of Bayesian inference for extended Skyrme, we had to thin the original chains less. 
If previously they were thinned up to the ``thinned'' autocorrelation length $\tau_\mathrm{th} = 1.0-1.2$, here we only thinned the chains up to $\tau_\mathrm{th} = 1.2-2.1$, thus reducing the number of independent samples in the posterior to $\gtrsim 48$\% of the number of samples (in the worst case).

Second, for run~0 the autocorrelation length estimate was unreliable due to some of the walkers getting stuck for a large number of chain steps. 
Thus, for this run we decided to switch to \textsc{ptemcee} sampler (v.1.0.0)~\cite{emcee,ptemcee}\,\footnote{The documentation is available at \url{https://emcee.readthedocs.io/en/v2.2.1/user/pt/} and the GitHub page is \url{https://github.com/willvousden/ptemcee}. Note that \textsc{ptemcee} is no longer maintained.} with tempered chains.
Specifically, we employed 5 chains with the maximum temperature $T = 512$ and auto-adjusting temperatures for intermediate chains.
The latter aimed to equalize the inter-chain swap probabilities, which were $\approx 30\%$. 
We made sure that even the highest temperature chain cannot ``break through'' the physical constraints on Fermi velocity and thermodynamic stability. 
With tempered chains the autocorrelation length is no longer an issue as each chain swap almost entirely erases all intra-chain correlations; in other words, $\tau$ does not exceed the number of steps between two consecutive chain swaps.
Nevertheless, even the highest temperature chain had low acceptance fraction ($\approx 15\%$) and advanced slowly, so we still had to thin the chains to compensate for this. 
With autocorrelation length no longer being a viable diagnostics, we used bootstrap analysis and the fraction of repetitions in the main chain as our diagnostic tools and to determine the necessary thinning factor. 
In particular, we thinned the chains until the fraction of repeated samples in the main chain dropped below $\lesssim 10^{-4}$.

In the same manner as in Paper~I, to additionally verify the correctness of the results, the posterior for the most numerically problematic run (run~0) was calculated independently by means of a nested sampling \citep{Skilling_2004,Skilling_2006} algorithm MLFriends~\cite{Buchner_2016,Buchner_2019}, implemented in the Python package \textsc{ultranest} (v.3.6.4)~\cite{ultranest}\,\footnote{The documentation is available at \url{https://johannesbuchner.github.io/UltraNest/} and the GitHub page is \url{https://github.com/JohannesBuchner/UltraNest}.}.
We had to introduce minor custom  adjustments to the algorithm responsible for the choice of initial positions of the live points. 
This was necessary to deal with likelihood plateaus occurring in the very beginning of the sampling as \textsc{ultranest} default algorithm was inefficient for our problem. 
No other modifications were done. 
After $\approx 1.1\times 10^{10}$ likelihood evaluations we ended up with $\approx 69\,000$ effective samples of the posterior (we employed inefficient, but robust region sampling). 
We have computed the values of 10\%, 50\% and 90\% quantiles and compared \textsc{ultranest} results against \textsc{ptemcee} results. 
The maximum difference between the corresponding values did not exceed 6\% and typical differences were less than 1\%.
Moreover, the maximum difference corresponded to a case where a quantile value was close to zero.
Plotted on a figure, the histograms were very close.

Thus, we have confirmed that despite above-mentioned difficulties our results are reliable. 

\section{Fermi velocity and effective mass constraints}
\label{App:vF}

\renewcommand{\arraystretch}{1.3}
\setlength{\tabcolsep}{3.7pt}
\begin{table*}  
	\caption{Median and 68\% CI of marginalized posterior distributions corresponding to key properties of NM and NSs. Reported data correspond to runs~0, 1, 1$^*$, 2, and 3.
		For the significance of each entry, see the text.
		The median values and the values of the CI are rounded to 3 and 2 significant digits, respectively; \emph{all} trailing zeros after the decimal point are removed.}
	\label{tab:Posteriors}
	\begin{tabular}{cccccccccccc}
		\toprule
		\toprule
		\multirow{2}{*}{Par.}               &  \multirow{2}{*}{Units}       & \multicolumn{2}{c}{run 0}                  & \multicolumn{2}{c}{run 1}                  & \multicolumn{2}{c}{run 1$^*$}              & \multicolumn{2}{c}{run 2}                  & \multicolumn{2}{c}{run 3}                  \\
		\cmidrule(lr){3-4}
		\cmidrule(lr){5-6}
		\cmidrule(lr){7-8}
		\cmidrule(lr){9-10}
		\cmidrule(lr){11-12}
		&                               &      Med.   &         68\% CI              &      Med.   &            68\% CI           &       Med.  &               68\% CI        &     Med.    &              68\% CI         &      Med.   &        68\% CI               \\ 
		\midrule
		
		$n_\mathrm{sat}$                    & $\mathrm{fm}^{-3}$            & $     0.16$ & $^{+   0.0038}_{-   0.0038}$ & $    0.161$ & $^{+   0.0038}_{-   0.0038}$ & $     0.16$ & $^{+   0.0039}_{-   0.0039}$ & $    0.161$ & $^{+   0.0038}_{-   0.0038}$ & $    0.161$ & $^{+   0.0038}_{-   0.0038}$ \\
		$E_\mathrm{sat}$                    & $\mathrm{MeV}$                & $    -15.9$ & $^{+      0.2}_{-      0.2}$ & $    -15.9$ & $^{+      0.2}_{-      0.2}$ & $    -15.9$ & $^{+      0.2}_{-      0.2}$ & $    -15.9$ & $^{+      0.2}_{-      0.2}$ & $    -15.9$ & $^{+      0.2}_{-      0.2}$ \\
		$K_\mathrm{sat}$                    & $\mathrm{MeV}$                & $      235$ & $^{+       12}_{-       12}$ & $      255$ & $^{+       18}_{-       18}$ & $      263$ & $^{+       23}_{-       21}$ & $      236$ & $^{+       14}_{-       13}$ & $      256$ & $^{+       10}_{-       10}$ \\
		$Q_\mathrm{sat}$                    & $\mathrm{MeV}$                & $     -434$ & $^{+       32}_{-       40}$ & $     -383$ & $^{+       61}_{-       55}$ & $     -364$ & $^{+       84}_{-       72}$ & $     -415$ & $^{+       36}_{-       31}$ & $     -388$ & $^{+       47}_{-       48}$ \\
		$Z_\mathrm{sat}$                    & $\mathrm{MeV}$                & $     1760$ & $^{+      280}_{-      290}$ & $     1250$ & $^{+      460}_{-      440}$ & $     1050$ & $^{+      520}_{-      520}$ & $     1710$ & $^{+      290}_{-      340}$ & $     1210$ & $^{+      270}_{-      240}$ \\
		$J_\mathrm{sym}$                    & $\mathrm{MeV}$                & $     29.8$ & $^{+     0.96}_{-     0.93}$ & $     29.9$ & $^{+     0.98}_{-     0.91}$ & $     30.1$ & $^{+      1.3}_{-      1.4}$ & $     30.9$ & $^{+      0.7}_{-     0.69}$ & $     31.2$ & $^{+     0.69}_{-     0.69}$ \\
		$L_\mathrm{sym}$                    & $\mathrm{MeV}$                & $     40.7$ & $^{+        5}_{-      4.9}$ & $     46.5$ & $^{+      7.1}_{-        7}$ & $     53.9$ & $^{+       11}_{-       10}$ & $     49.8$ & $^{+      5.7}_{-      5.7}$ & $     52.7$ & $^{+      5.6}_{-      6.2}$ \\
		$K_\mathrm{sym}$                    & $\mathrm{MeV}$                & $     -146$ & $^{+       18}_{-       18}$ & $     -122$ & $^{+       33}_{-       28}$ & $    -29.5$ & $^{+      110}_{-       77}$ & $     -131$ & $^{+       15}_{-       16}$ & $     -135$ & $^{+       17}_{-       20}$ \\
		$Q_\mathrm{sym}$                    & $\mathrm{MeV}$                & $      629$ & $^{+       86}_{-       69}$ & $      590$ & $^{+      120}_{-      120}$ & $      722$ & $^{+      360}_{-      220}$ & $      507$ & $^{+       57}_{-       67}$ & $      435$ & $^{+       65}_{-       69}$ \\
		$Z_\mathrm{sym}$                    & $\mathrm{MeV}$                & $    -2460$ & $^{+      380}_{-      320}$ & $    -2420$ & $^{+      480}_{-      430}$ & $    -3440$ & $^{+      890}_{-     1200}$ & $    -2170$ & $^{+      270}_{-      260}$ & $    -1780$ & $^{+      170}_{-      160}$ \\
		$m_{\mathrm{eff;\,N}}^\mathrm{SNM}$ & $m_{\mathrm{N}}$              & $    0.532$ & $^{+    0.092}_{-    0.063}$ & $    0.539$ & $^{+    0.096}_{-    0.061}$ & $    0.265$ & $^{+     0.13}_{-    0.077}$ & $    0.634$ & $^{+   0.0086}_{-   0.0089}$ & $    0.626$ & $^{+   0.0083}_{-   0.0083}$ \\
		$(E/A)_1$                           & $\mathrm{MeV}$                & $     6.22$ & $^{+     0.06}_{-     0.06}$ & $     6.22$ & $^{+     0.25}_{-     0.25}$ & $     6.89$ & $^{+     0.64}_{-     0.52}$ & $     5.87$ & $^{+      0.2}_{-     0.21}$ & $     5.92$ & $^{+     0.17}_{-     0.18}$ \\
		$(E/A)_2$                           & $\mathrm{MeV}$                & $     9.32$ & $^{+      0.2}_{-      0.2}$ & $     9.14$ & $^{+     0.22}_{-     0.21}$ & $     9.23$ & $^{+     0.22}_{-     0.22}$ & $     9.02$ & $^{+      0.2}_{-      0.2}$ & $     8.99$ & $^{+      0.2}_{-      0.2}$ \\
		$(E/A)_3$                           & $\mathrm{MeV}$                & $     12.6$ & $^{+     0.45}_{-     0.44}$ & $     12.4$ & $^{+     0.28}_{-     0.28}$ & $     12.1$ & $^{+     0.35}_{-     0.36}$ & $     12.6$ & $^{+     0.27}_{-     0.27}$ & $     12.6$ & $^{+     0.27}_{-     0.27}$ \\
		$(E/A)_4$                           & $\mathrm{MeV}$                & $     16.4$ & $^{+     0.77}_{-     0.75}$ & $     16.5$ & $^{+     0.64}_{-     0.64}$ & $     16.4$ & $^{+      0.7}_{-      0.7}$ & $     16.8$ & $^{+     0.58}_{-     0.58}$ & $       17$ & $^{+     0.56}_{-     0.57}$ \\
		$m_{\mathrm{eff;\,n}}^\mathrm{PNM}$ & $m_{\mathrm{n}}$              & $    0.848$ & $^{+    0.079}_{-    0.076}$ & $    0.892$ & $^{+    0.062}_{-    0.079}$ & $    0.248$ & $^{+     0.19}_{-    0.087}$ & $    0.852$ & $^{+    0.011}_{-    0.011}$ & $    0.823$ & $^{+   0.0071}_{-    0.007}$ \\
		$M_\mathrm{G}^*$                    & $\Msun$                       & $     2.11$ & $^{+    0.048}_{-    0.034}$ & $     2.13$ & $^{+    0.086}_{-    0.059}$ & $     2.18$ & $^{+     0.12}_{-      0.1}$ & $     2.08$ & $^{+    0.039}_{-    0.023}$ & $     2.09$ & $^{+    0.046}_{-    0.029}$ \\
		$c^{*2}_\mathrm{s}$                 & $c^2$                         & $    0.928$ & $^{+    0.051}_{-    0.089}$ & $    0.885$ & $^{+    0.083}_{-     0.17}$ & $    0.789$ & $^{+     0.15}_{-     0.21}$ & $    0.969$ & $^{+    0.023}_{-    0.041}$ & $    0.945$ & $^{+    0.038}_{-    0.057}$ \\
		$n_\mathrm{c}^*$                    & $\mathrm{fm}^{-3}$            & $     1.15$ & $^{+    0.041}_{-    0.042}$ & $      1.1$ & $^{+    0.059}_{-     0.07}$ & $    0.991$ & $^{+    0.086}_{-    0.073}$ & $     1.18$ & $^{+    0.031}_{-    0.045}$ & $     1.17$ & $^{+    0.038}_{-    0.049}$ \\
		$\rho_\mathrm{c,15}^*$              & $10^{15}~\mathrm{g/cm^{3}}$   & $     2.71$ & $^{+     0.12}_{-     0.11}$ & $     2.56$ & $^{+     0.17}_{-     0.17}$ & $     2.27$ & $^{+     0.22}_{-     0.18}$ & $      2.8$ & $^{+    0.083}_{-     0.11}$ & $     2.76$ & $^{+     0.11}_{-     0.12}$ \\
		$P_\mathrm{c,36}^*$                 & $\mathrm{10^{36}~dyn/cm^{2}}$ & $     1.32$ & $^{+     0.12}_{-     0.14}$ & $     1.17$ & $^{+     0.15}_{-     0.18}$ & $    0.922$ & $^{+     0.17}_{-     0.19}$ & $     1.38$ & $^{+    0.085}_{-    0.085}$ & $     1.32$ & $^{+     0.13}_{-    0.092}$ \\
		$R_{1.4}$                           & $\mathrm{km}$                 & $     11.6$ & $^{+     0.21}_{-     0.22}$ & $     12.1$ & $^{+     0.32}_{-     0.34}$ & $     12.7$ & $^{+      0.5}_{-     0.49}$ & $     11.8$ & $^{+     0.28}_{-     0.27}$ & $       12$ & $^{+     0.27}_{-     0.34}$ \\
		$\Lambda_{1.4}$                     & --                            & $      292$ & $^{+       37}_{-       35}$ & $      372$ & $^{+       73}_{-       64}$ & $      539$ & $^{+      160}_{-      130}$ & $      306$ & $^{+       51}_{-       42}$ & $      338$ & $^{+       54}_{-       57}$ \\
		$R_{2.0}$                           & $\mathrm{km}$                 & $       11$ & $^{+     0.29}_{-     0.27}$ & $     11.4$ & $^{+      0.5}_{-     0.42}$ & $     12.2$ & $^{+     0.65}_{-     0.69}$ & $     10.9$ & $^{+     0.36}_{-     0.25}$ & $       11$ & $^{+     0.39}_{-     0.32}$ \\
		$\Lambda_{2.0}$                     & --                            & $     15.3$ & $^{+      4.1}_{-      3.1}$ & $       21$ & $^{+      9.6}_{-        6}$ & $     38.1$ & $^{+       20}_{-       15}$ & $     13.5$ & $^{+      4.3}_{-      2.5}$ & $       15$ & $^{+      5.4}_{-      3.3}$ \\    
		\bottomrule
		\bottomrule
	\end{tabular}
\end{table*}
\renewcommand{\arraystretch}{1.0}
\setlength{\tabcolsep}{2.0pt}

Here, we investigate the sensitivity of our results to the density up to which the constraints on
the Fermi velocity and effective masses are imposed (hereafter, limiting density).
To this end, two runs equivalent to run~1 but with these constraints imposed up to 0.5~fm$^{-3}$ (run~1a) and 1.0~fm$^{-3}$ (run~1b) are considered.
Note that, since for non-relativistic models $v_\mathrm{F} = p_\mathrm{F}/m$, the constraint from above on $v_\mathrm{F}$ is, essentially, a density-dependent constraint from below on the effective mass. 
This is the reason why for runs~1a and 1b we vary the limiting density not only for the Fermi velocity constraint, but for the effective masses constraints as well.

Fig.~\ref{Fig:Hist_NM_vF} is the analogue of Fig.~\ref{Fig:Hist_NM} and shows the marginalized posterior distributions of NEPs, effective masses, and $E/A$ in PNM. 
One can see that the impact of the limiting density depends on the quantity under consideration.
The quantities that are tightly constrained (e.g., $\sat{n}, \sat{E}$) are not affected, those that are loosely constrained (e.g., $\sym{J}$) are affected slightly. Finally, the quantities that are practically unconstrained (e.g., $\sym{K}$, $\sym{Q}$, $\sym{Z}$) manifest the largest modifications. 
In general, the changes are never too large and run~1b has a tendency to provide narrower distributions.
The latter fact is not surprising, as run~1b has the strongest constraints among the three considered runs.

Fig.~\ref{Fig:Hist_NS_vF} is the analogue of Fig.~\ref{Fig:Hist_NS} and shows the marginalized posterior distributions of some selected properties of NSs. 
In addition to results of runs~1, 1a, and 1b, considered in Fig.~\ref{Fig:Hist_NM_vF}, here we show also results of the run~1$^*$, which has no $v_\mathrm{n;\,F}$ constraint, as well as those of run~1 from Paper~I and run~4 from Ref.~\cite{Beznogov_PRC_2023}. 
First, we note that increasing the limiting density does not suffice to make the NS EOS physical over the whole density domain spanned in massive NSs. 
Then, it is obvious that the value chosen for the limiting density impacts all NS properties. 
As expected based on the results in Fig.~\ref{Fig:Hist_NS},
none of the posteriors of runs~1a, 1b resemble the posteriors of run~1 in Paper I. In exchange, it appears that the posteriors of $n_{\mathrm c}^*$, $R_{2.0}$, $\Lambda_{2.0}$ for run~1$^*$ here and run~4 in Ref.~\cite{Beznogov_PRC_2023} are similar. We consider this as accidental.

\section{Posterior data table}
\label{App:TableNMandNS}

In Table~\ref{tab:Posteriors} we provide information complementary to Figs.~\ref{Fig:Hist_NM} and \ref{Fig:Hist_NS}.

\bibliography{BSk.bib,MCMC.bib}
\end{document}